\documentclass[10pt,twoside,twocolumn]{IEEEtran}

\usepackage{cite}
\usepackage[T1]{fontenc}
\usepackage{graphicx}
\usepackage{amssymb}
\usepackage{amsmath}
\usepackage{amsthm}
\usepackage{subfigure}
\usepackage{booktabs} 
\usepackage{multirow}
\usepackage{microtype}
\usepackage{balance}
\usepackage{xcolor}
\usepackage{xfrac}    
\usepackage{algorithm}
\usepackage{setspace}
\usepackage{mdframed}
\usepackage{tikz}
\usepackage{circledsteps}
\usepackage{enumitem,kantlipsum}
\usepackage{twemojis}

\usepackage{algorithmicx}
\usepackage{algpseudocode}
\algrenewcommand\algorithmicindent{0.7em}%
\usepackage{xpatch}

\usepackage[hidelinks]{hyperref}

\usepackage{amssymb}
\usepackage{amsfonts}
\usepackage{mathrsfs}
\usepackage{xspace}
\usepackage{bm}
\usepackage{upgreek}

\newcommand{\safemath}[2]{\newcommand{#1}{\ensuremath{#2}\xspace}}

\safemath{\bma}{\mathbf{a}}
\safemath{\bmb}{\mathbf{b}}
\safemath{\bmc}{\mathbf{c}}
\safemath{\bmd}{\mathbf{d}}
\safemath{\bme}{\mathbf{e}}
\safemath{\bmf}{\mathbf{f}}
\safemath{\bmg}{\mathbf{g}}
\safemath{\bmh}{\mathbf{h}}
\safemath{\bmi}{\mathbf{i}}
\safemath{\bmj}{\mathbf{j}}
\safemath{\bmk}{\mathbf{k}}
\safemath{\bml}{\mathbf{l}}
\safemath{\bmm}{\mathbf{m}}
\safemath{\bmn}{\mathbf{n}}
\safemath{\bmo}{\mathbf{o}}
\safemath{\bmp}{\mathbf{p}}
\safemath{\bmq}{\mathbf{q}}
\safemath{\bmr}{\mathbf{r}}
\safemath{\bms}{\mathbf{s}}
\safemath{\bmt}{\mathbf{t}}
\safemath{\bmu}{\mathbf{u}}
\safemath{\bmv}{\mathbf{v}}
\safemath{\bmw}{\mathbf{w}}
\safemath{\bmx}{\mathbf{x}}
\safemath{\bmy}{\mathbf{y}}
\safemath{\bmz}{\mathbf{z}}
\safemath{\bmzero}{\mathbf{0}}
\safemath{\bmone}{\mathbf{1}}

\bmdefine{\biad}{a}
\bmdefine{\bibd}{b}
\bmdefine{\bicd}{c}
\bmdefine{\bidd}{d}
\bmdefine{\bied}{e}
\bmdefine{\bifd}{f}
\bmdefine{\bigd}{g}
\bmdefine{\bihd}{h}
\bmdefine{\biid}{i}
\bmdefine{\bijd}{j}
\bmdefine{\bikd}{k}
\bmdefine{\bild}{l}
\bmdefine{\bimd}{m}
\bmdefine{\bind}{n}
\bmdefine{\biod}{o}
\bmdefine{\bipd}{p}
\bmdefine{\biqd}{q}
\bmdefine{\bird}{r}
\bmdefine{\bisd}{s}
\bmdefine{\bitd}{t}
\bmdefine{\biud}{u}
\bmdefine{\bivd}{v}
\bmdefine{\biwd}{w}
\bmdefine{\bixd}{x}
\bmdefine{\biyd}{y}
\bmdefine{\bizd}{z}

\bmdefine{\bixid}{\xi}
\bmdefine{\bilambdad}{\lambda}
\bmdefine{\bimud}{\mu}
\bmdefine{\bithetad}{\theta}
\bmdefine{\biphid}{\phi}
\bmdefine{\bideltad}{\delta}

\safemath{\bmia}{\biad}
\safemath{\bmib}{\bibd}
\safemath{\bmic}{\bicd}
\safemath{\bmid}{\bidd}
\safemath{\bmie}{\bied}
\safemath{\bmif}{\bifd}
\safemath{\bmig}{\bigd}
\safemath{\bmih}{\bihd}
\safemath{\bmii}{\biid}
\safemath{\bmij}{\bijd}
\safemath{\bmik}{\bikd}
\safemath{\bmil}{\bild}
\safemath{\bmim}{\bimd}
\safemath{\bmin}{\bind}
\safemath{\bmio}{\biod}
\safemath{\bmip}{\bipd}
\safemath{\bmiq}{\biqd}
\safemath{\bmir}{\bird}
\safemath{\bmis}{\bisd}
\safemath{\bmit}{\bitd}
\safemath{\bmiu}{\biud}
\safemath{\bmiv}{\bivd}
\safemath{\bmiw}{\biwd}
\safemath{\bmix}{\bixd}
\safemath{\bmiy}{\biyd}
\safemath{\bmiz}{\bizd}

\safemath{\bmxi}{\bixid}
\safemath{\bmlambda}{\bilambdad}
\safemath{\bmmu}{\bimud}
\safemath{\bmtheta}{\bithetad}
\safemath{\bmphi}{\biphid}
\safemath{\bmdelta}{\bideltad}

\safemath{\bA}{\mathbf{A}}
\safemath{\bB}{\mathbf{B}}
\safemath{\bC}{\mathbf{C}}
\safemath{\bD}{\mathbf{D}}
\safemath{\bE}{\mathbf{E}}
\safemath{\bF}{\mathbf{F}}
\safemath{\bG}{\mathbf{G}}
\safemath{\bH}{\mathbf{H}}
\safemath{\bI}{\mathbf{I}}
\safemath{\bJ}{\mathbf{J}}
\safemath{\bK}{\mathbf{K}}
\safemath{\bL}{\mathbf{L}}
\safemath{\bM}{\mathbf{M}}
\safemath{\bN}{\mathbf{N}}
\safemath{\bO}{\mathbf{O}}
\safemath{\bP}{\mathbf{P}}
\safemath{\bQ}{\mathbf{Q}}
\safemath{\bR}{\mathbf{R}}
\safemath{\bS}{\mathbf{S}}
\safemath{\bT}{\mathbf{T}}
\safemath{\bU}{\mathbf{U}}
\safemath{\bV}{\mathbf{V}}
\safemath{\bW}{\mathbf{W}}
\safemath{\bX}{\mathbf{X}}
\safemath{\bY}{\mathbf{Y}}
\safemath{\bZ}{\mathbf{Z}}

\safemath{\bZero}{\mathbf{0}}
\safemath{\bOne}{\mathbf{1}}
\safemath{\bDelta}{\mathbf{\Delta}}
\safemath{\bLambda}{\mathbf{\UpLambda}}
\safemath{\bPhi}{\mathbf{\Upphi}}
\safemath{\bSigma}{\mathbf{\Upsigma}}
\safemath{\bOmega}{\mathbf{\Upomega}}
\safemath{\bTheta}{\mathbf{\Uptheta}}

\bmdefine{\biAd}{A}
\bmdefine{\biBd}{B}
\bmdefine{\biCd}{C}
\bmdefine{\biDd}{D}
\bmdefine{\biEd}{E}
\bmdefine{\biFd}{F}
\bmdefine{\biGd}{G}
\bmdefine{\biHd}{H}
\bmdefine{\biId}{I}
\bmdefine{\biJd}{J}
\bmdefine{\biKd}{K}
\bmdefine{\biLd}{L}
\bmdefine{\biMd}{M}
\bmdefine{\biOd}{N}
\bmdefine{\biPd}{O}
\bmdefine{\biQd}{P}
\bmdefine{\biRd}{R}
\bmdefine{\biSd}{S}
\bmdefine{\biTd}{T}
\bmdefine{\biUd}{U}
\bmdefine{\biVd}{V}
\bmdefine{\biWd}{W}
\bmdefine{\biXd}{X}
\bmdefine{\biYd}{Y}
\bmdefine{\biZd}{Z}

\bmdefine{\biDelta}{\Delta}
\bmdefine{\biLambda}{\Lambda}
\bmdefine{\biPhi}{\Phi}
\bmdefine{\biSigma}{\Sigma}
\bmdefine{\biOmega}{\Omega}
\bmdefine{\biTheta}{\Theta}

\safemath{\bimA}{\biAd}
\safemath{\bimB}{\biBd}
\safemath{\bimC}{\biCd}
\safemath{\bimD}{\biDd}
\safemath{\bimE}{\biEd}
\safemath{\bimF}{\biFd}
\safemath{\bimG}{\biGd}
\safemath{\bimH}{\biHd}
\safemath{\bimI}{\biId}
\safemath{\bimJ}{\biJd}
\safemath{\bimK}{\biKd}
\safemath{\bimL}{\biLd}
\safemath{\bimM}{\biMd}
\safemath{\bimN}{\biNd}
\safemath{\bimO}{\biOd}
\safemath{\bimP}{\biPd}
\safemath{\bimQ}{\biQd}
\safemath{\bimR}{\biRd}
\safemath{\bimS}{\biSd}
\safemath{\bimT}{\biTd}
\safemath{\bimU}{\biUd}
\safemath{\bimV}{\biVd}
\safemath{\bimW}{\biWd}
\safemath{\bimX}{\biXd}
\safemath{\bimY}{\biYd}
\safemath{\bimZ}{\biZd}

\safemath{\bimDelta}{\biDelta}
\safemath{\bimLambda}{\biLambda}
\safemath{\bimPhi}{\biPhi}
\safemath{\bimSigma}{\biSigma}
\safemath{\bimOmega}{\biOmega}
\safemath{\bimTheta}{\biTheta}

\safemath{\setA}{\mathcal{A}}
\safemath{\setB}{\mathcal{B}}
\safemath{\setC}{\mathcal{C}}
\safemath{\setD}{\mathcal{D}}
\safemath{\setE}{\mathcal{E}}
\safemath{\setF}{\mathcal{F}}
\safemath{\setG}{\mathcal{G}}
\safemath{\setH}{\mathcal{H}}
\safemath{\setI}{\mathcal{I}}
\safemath{\setJ}{\mathcal{J}}
\safemath{\setK}{\mathcal{K}}
\safemath{\setL}{\mathcal{L}}
\safemath{\setM}{\mathcal{M}}
\safemath{\setN}{\mathcal{N}}
\safemath{\setO}{\mathcal{O}}
\safemath{\setP}{\mathcal{P}}
\safemath{\setQ}{\mathcal{Q}}
\safemath{\setR}{\mathcal{R}}
\safemath{\setS}{\mathcal{S}}
\safemath{\setT}{\mathcal{T}}
\safemath{\setU}{\mathcal{U}}
\safemath{\setV}{\mathcal{V}}
\safemath{\setW}{\mathcal{W}}
\safemath{\setX}{\mathcal{X}}
\safemath{\setY}{\mathcal{Y}}
\safemath{\setZ}{\mathcal{Z}}
\safemath{\emptySet}{\varnothing}

\safemath{\colA}{\mathscr{A}}
\safemath{\colB}{\mathscr{B}}
\safemath{\colC}{\mathscr{C}}
\safemath{\colD}{\mathscr{D}}
\safemath{\colE}{\mathscr{E}}
\safemath{\colF}{\mathscr{F}}
\safemath{\colG}{\mathscr{G}}
\safemath{\colH}{\mathscr{H}}
\safemath{\colI}{\mathscr{I}}
\safemath{\colJ}{\mathscr{J}}
\safemath{\colK}{\mathscr{K}}
\safemath{\colL}{\mathscr{L}}
\safemath{\colM}{\mathscr{M}}
\safemath{\colN}{\mathscr{N}}
\safemath{\colO}{\mathscr{O}}
\safemath{\colP}{\mathscr{P}}
\safemath{\colQ}{\mathscr{Q}}
\safemath{\colR}{\mathscr{R}}
\safemath{\colS}{\mathscr{S}}
\safemath{\colT}{\mathscr{T}}
\safemath{\colU}{\mathscr{U}}
\safemath{\colV}{\mathscr{V}}
\safemath{\colW}{\mathscr{W}}
\safemath{\colX}{\mathscr{X}}
\safemath{\colY}{\mathscr{Y}}
\safemath{\colZ}{\mathscr{Z}}

\safemath{\opA}{\mathbb{A}}
\safemath{\opB}{\mathbb{B}}
\safemath{\opC}{\mathbb{C}}
\safemath{\opD}{\mathbb{D}}
\safemath{\opE}{\mathbb{E}}
\safemath{\opF}{\mathbb{F}}
\safemath{\opG}{\mathbb{G}}
\safemath{\opH}{\mathbb{H}}
\safemath{\opI}{\mathbb{I}}
\safemath{\opJ}{\mathbb{J}}
\safemath{\opK}{\mathbb{K}}
\safemath{\opL}{\mathbb{L}}
\safemath{\opM}{\mathbb{M}}
\safemath{\opN}{\mathbb{N}}
\safemath{\opO}{\mathbb{O}}
\safemath{\opP}{\mathbb{P}}
\safemath{\opQ}{\mathbb{Q}}
\safemath{\opR}{\mathbb{R}}
\safemath{\opS}{\mathbb{S}}
\safemath{\opT}{\mathbb{T}}
\safemath{\opU}{\mathbb{U}}
\safemath{\opV}{\mathbb{V}}
\safemath{\opW}{\mathbb{W}}
\safemath{\opX}{\mathbb{X}}
\safemath{\opY}{\mathbb{Y}}
\safemath{\opZ}{\mathbb{Z}}
\safemath{\opZero}{\mathbb{O}}
\safemath{\identityop}{\opI}

\safemath{\veca}{\bma}
\safemath{\vecb}{\bmb}
\safemath{\vecc}{\bmc}
\safemath{\vecd}{\bmd}
\safemath{\vece}{\bme}
\safemath{\vecf}{\bmf}
\safemath{\vecg}{\bmg}
\safemath{\vech}{\bmh}
\safemath{\veci}{\bmi}
\safemath{\vecj}{\bmj}
\safemath{\veck}{\bmk}
\safemath{\vecl}{\bml}
\safemath{\vecm}{\bmm}
\safemath{\vecn}{\bmn}
\safemath{\veco}{\bmo}
\safemath{\vecp}{\bmp}
\safemath{\vecq}{\bmq}
\safemath{\vecr}{\bmr}
\safemath{\vecs}{\bms}
\safemath{\vect}{\bmt}
\safemath{\vecu}{\bmu}
\safemath{\vecv}{\bmv}
\safemath{\vecw}{\bmw}
\safemath{\vecx}{\bmx}
\safemath{\vecy}{\bmy}
\safemath{\vecz}{\bmz}

\safemath{\veczero}{\bmzero}
\safemath{\vecone}{\bmone}
\safemath{\vecxi}{\bmxi}
\safemath{\veclambda}{\bmlambda}
\safemath{\vecmu}{\bmmu}
\safemath{\vectheta}{\bmtheta}
\safemath{\vecphi}{\bmphi}
\safemath{\vecdelta}{\bmdelta}

\safemath{\matA}{\bA}
\safemath{\matB}{\bB}
\safemath{\matC}{\bC}
\safemath{\matD}{\bD}
\safemath{\matE}{\bE}
\safemath{\matF}{\bF}
\safemath{\matG}{\bG}
\safemath{\matH}{\bH}
\safemath{\matI}{\bI}
\safemath{\matJ}{\bJ}
\safemath{\matK}{\bK}
\safemath{\matL}{\bL}
\safemath{\matM}{\bM}
\safemath{\matN}{\bN}
\safemath{\matO}{\bO}
\safemath{\matP}{\bP}
\safemath{\matQ}{\bQ}
\safemath{\matR}{\bR}
\safemath{\matS}{\bS}
\safemath{\matT}{\bT}
\safemath{\matU}{\bU}
\safemath{\matV}{\bV}
\safemath{\matW}{\bW}
\safemath{\matX}{\bX}
\safemath{\matY}{\bY}
\safemath{\matZ}{\bZ}
\safemath{\matzero}{\bmzero}

\safemath{\matDelta}{\bDelta}
\safemath{\matLambda}{\bLambda}
\safemath{\matPhi}{\bPhi}
\safemath{\matSigma}{\bSigma}
\safemath{\matOmega}{\bOmega}
\safemath{\matTheta}{\bTheta}

\safemath{\matidentity}{\matI}
\safemath{\matone}{\matO}

\safemath{\rnda}{A}
\safemath{\rndb}{B}
\safemath{\rndc}{C}
\safemath{\rndd}{D}
\safemath{\rnde}{E}
\safemath{\rndf}{F}
\safemath{\rndg}{G}
\safemath{\rndh}{H}
\safemath{\rndi}{I}
\safemath{\rndj}{J}
\safemath{\rndk}{K}
\safemath{\rndl}{L}
\safemath{\rndm}{M}
\safemath{\rndn}{N}
\safemath{\rndo}{O}
\safemath{\rndp}{P}
\safemath{\rndq}{Q}
\safemath{\rndr}{R}
\safemath{\rnds}{S}
\safemath{\rndt}{T}
\safemath{\rndu}{U}
\safemath{\rndv}{V}
\safemath{\rndw}{W}
\safemath{\rndx}{X}
\safemath{\rndy}{Y}
\safemath{\rndz}{Z}

\safemath{\rveca}{\bimA}
\safemath{\rvecb}{\bimB}
\safemath{\rvecc}{\bimC}
\safemath{\rvecd}{\bimD}
\safemath{\rvece}{\bimE}
\safemath{\rvecf}{\bimF}
\safemath{\rvecg}{\bimG}
\safemath{\rvech}{\bimH}
\safemath{\rveci}{\bimI}
\safemath{\rvecj}{\bimJ}
\safemath{\rveck}{\bimK}
\safemath{\rvecl}{\bimL}
\safemath{\rvecm}{\bimM}
\safemath{\rvecn}{\bimN}
\safemath{\rveco}{\bomO}
\safemath{\rvecp}{\bimP}
\safemath{\rvecq}{\bimQ}
\safemath{\rvecr}{\bimR}
\safemath{\rvecs}{\bimS}
\safemath{\rvect}{\bimT}
\safemath{\rvecu}{\bimU}
\safemath{\rvecv}{\bimV}
\safemath{\rvecw}{\bimW}
\safemath{\rvecx}{\bimX}
\safemath{\rvecy}{\bimY}
\safemath{\rvecz}{\bimZ}

\safemath{\rvecxi}{\bmxi}
\safemath{\rveclambda}{\bmlambda}
\safemath{\rvecmu}{\bmmu}
\safemath{\rvectheta}{\bmtheta}
\safemath{\rvecphi}{\bmphi}

\safemath{\rmatA}{\bimA}
\safemath{\rmatB}{\bimB}
\safemath{\rmatC}{\bimC}
\safemath{\rmatD}{\bimD}
\safemath{\rmatE}{\bimE}
\safemath{\rmatF}{\bimF}
\safemath{\rmatG}{\bimG}
\safemath{\rmatH}{\bimH}
\safemath{\rmatI}{\bimI}
\safemath{\rmatJ}{\bimJ}
\safemath{\rmatK}{\bimK}
\safemath{\rmatL}{\bimL}
\safemath{\rmatM}{\bimM}
\safemath{\rmatN}{\bimN}
\safemath{\rmatO}{\bimO}
\safemath{\rmatP}{\bimP}
\safemath{\rmatQ}{\bimQ}
\safemath{\rmatR}{\bimR}
\safemath{\rmatS}{\bimS}
\safemath{\rmatT}{\bimT}
\safemath{\rmatU}{\bimU}
\safemath{\rmatV}{\bimV}
\safemath{\rmatW}{\bimW}
\safemath{\rmatX}{\bimX}
\safemath{\rmatY}{\bimY}
\safemath{\rmatZ}{\bimZ}

\safemath{\rmatDelta}{\bimDelta}
\safemath{\rmatLambda}{\bimLambda}
\safemath{\rmatPhi}{\bimPhi}
\safemath{\rmatSigma}{\bimSigma}
\safemath{\rmatOmega}{\bimOmega}
\safemath{\rmatTheta}{\bimTheta}

\usepackage{amssymb}
\usepackage{amsfonts}
\usepackage{mathrsfs}
\usepackage{xspace}
\usepackage{bm}
\usepackage{fancyref}
\usepackage{textcomp}

\usepackage{multirow}
\usepackage{stmaryrd}

\newenvironment{textbmatrix}{	\setlength{\arraycolsep}{2.5pt}%
								\big[\begin{matrix}}{\end{matrix}\big]%
								\raisebox{0.08ex}{\vphantom{M}}}

\def\be{\begin{equation}}
\def\ee{\end{equation}}
\def\een{\nonumber \end{equation}}
\def\mat{\begin{bmatrix}}
\def\emat{\end{bmatrix}}
\def\btm{\begin{textbmatrix}}
\def\etm{\end{textbmatrix}}

\def\ba#1\ea{\begin{align}#1\end{align}}
\def\bas#1\eas{\begin{align*}#1\end{align*}}
\def\bs#1\es{\begin{split}#1\end{split}}
\def\bg#1\eg{\begin{gather}#1\end{gather}}
\def\bml#1\eml{\begin{multline}#1\end{multline}}
\def\bi#1\ei{\begin{itemize}#1\end{itemize}}

\newcommand{\lefto}{\mathopen{}\left}

\DeclareMathOperator*{\argmax}{arg\;max}		
\DeclareMathOperator{\Exop}{\opE}			

\newcommand{\orth}{\perp}					
\newcommand{\Ex}[2]{\ensuremath{\Exop_{#1}\lefto[#2\right]}} 	

\newcommand{\tp}[1]{\ensuremath{#1^{\text{T}}}} 		
\newcommand{\herm}[1]{\ensuremath{#1^{\text{H}}}} 	
\newcommand{\inv}[1]{\ensuremath{#1^{-1}}} 	
\newcommand{\pinv}[1]{\ensuremath{#1^{\dagger}}} 	

\safemath{\dirac}{\delta}					
\safemath{\krond}{\dirac}					

\safemath{\upto}{\uparrow}
\safemath{\downto}{\downarrow}
\safemath{\iu}{j}							
\safemath{\ev}{\lambda}						
\safemath{\hilseqspace}{l^{2}}				
\newcommand{\banachfunspace}[1]{\setL^{#1}}	
\safemath{\hilfunspace}{\banachfunspace{2}}	

\safemath{\SNR}{\textit{SNR}} 				
\safemath{\PAR}{\textit{PAR}} 				
\safemath{\No}{N_0}							
\safemath{\Es}{E_s}							
\safemath{\Eb}{E_b}							
\safemath{\EbNo}{\frac{\Eb}{\No}}
\safemath{\EsNo}{\frac{\Es}{\No}}

\DeclareMathOperator{\CHop}{\ensuremath{\opH}} 
\safemath{\tvir}{\rndh_{\CHop}}				
\safemath{\tvtf}{\rndl_{\CHop}}				
\safemath{\spf}{\rnds_{\CHop}}				
\safemath{\bff}{H_{\CHop}}					

\safemath{\ircf}{r_{h}}						
\safemath{\tftvcf}{r_{s}}					
\safemath{\tfcf}{r_{l}}						
\safemath{\bfcf}{r_{H}}						

\safemath{\tcorr}{c_h}						
\safemath{\scf}{c_{s}}						
\safemath{\tfcorr}{c_{l}}					
\safemath{\fcorr}{c_{H}}						

\safemath{\mi}{I}							
\safemath{\capacity}{C}						

\safemath{\normal}{\mathcal{N}}			
\safemath{\jpg}{\mathcal{CN}}			
\safemath{\mchain}{\leftrightarrow}		

\safemath{\dB}{\,\mathrm{dB}}
\safemath{\dBm}{\,\mathrm{dBm}}
\safemath{\Hz}{\,\mathrm{Hz}}
\safemath{\kHz}{\,\mathrm{kHz}}
\safemath{\MHz}{\,\mathrm{MHz}}
\safemath{\GHz}{\,\mathrm{GHz}}
\safemath{\s}{\,\mathrm{s}}
\safemath{\ms}{\,\mathrm{ms}}
\safemath{\mus}{\,\mathrm{\text{\textmu}s}}
\safemath{\ns}{\,\mathrm{ns}}
\safemath{\ps}{\,\mathrm{ps}}
\safemath{\meter}{\,\mathrm{m}}
\safemath{\mm}{\,\mathrm{mm}}
\safemath{\cm}{\,\mathrm{cm}}
\safemath{\m}{\,\mathrm{m}}
\safemath{\W}{\,\mathrm{W}}
\safemath{\mW}{\, \mathrm{mW}}
\safemath{\J}{\,\mathrm{J}}
\safemath{\K}{\,\mathrm{K}}
\safemath{\bit}{\,\mathrm{bit}}
\safemath{\nat}{\,\mathrm{nat}}

\safemath{\define}{\triangleq}			

\safemath{\equivalent}{\sim}
\safemath{\distas}{\sim}					
\safemath{\sdiff}{\Delta}				

\safemath{\reals}{\mathbb{R}}
\safemath{\positivereals}{\reals_{+}}
\safemath{\integers}{\mathbb{Z}}
\safemath{\posint}{\integers_{+}}
\safemath{\naturals}{\mathbb{N}}
\safemath{\posnaturals}{\naturals_{+}}
\safemath{\complexset}{\mathbb{C}}
\safemath{\rationals}{\mathbb{Q}}

\newcommand*{\fancyrefapplabelprefix}{app}		
\newcommand*{\fancyrefthmlabelprefix}{thm}		
\newcommand*{\fancyreflemlabelprefix}{lem}		
\newcommand*{\fancyrefcorlabelprefix}{cor}		
\newcommand*{\fancyrefdeflabelprefix}{def}		
\newcommand*{\fancyrefproplabelprefix}{prop}		
\newcommand*{\fancyrefexmpllabelprefix}{exmpl}
\newcommand*{\fancyrefalglabelprefix}{alg}		
\newcommand*{\fancyreftbllabelprefix}{tbl}		

\frefformat{vario}{\fancyrefseclabelprefix}{Section~#1}
\frefformat{vario}{\fancyrefthmlabelprefix}{Theorem~#1}
\frefformat{vario}{\fancyreftbllabelprefix}{Table~#1}
\frefformat{vario}{\fancyreflemlabelprefix}{Lemma~#1}
\frefformat{vario}{\fancyrefcorlabelprefix}{Corollary~#1}
\frefformat{vario}{\fancyrefdeflabelprefix}{Definition~#1}
\frefformat{vario}{\fancyreffiglabelprefix}{Fig.~#1}
\frefformat{vario}{\fancyrefapplabelprefix}{Appendix~#1}
\frefformat{vario}{\fancyrefeqlabelprefix}{(#1)}
\frefformat{vario}{\fancyrefproplabelprefix}{Proposition~#1}
\frefformat{vario}{\fancyrefexmpllabelprefix}{Example~#1}
\frefformat{vario}{\fancyrefalglabelprefix}{Algorithm~#1}

 \newtheorem{thm}{Theorem}
 \newtheorem{prop}{Proposition}
 \newtheorem{defi}{Definition}

 \newtheorem*{remark*}{Remark}

\safemath{\dictab}{[\,\dicta\,\,\dictb\,]}

\safemath{\ysig}{\bmy}
\safemath{\ysighat}{\hat{\ysig}}
\safemath{\ysigdim}{M}
\safemath{\xsig}{\bmx}
\safemath{\xsigdim}{N}
\safemath{\nx}{n_x}
\safemath{\zsig}{\bmz}
\safemath{\zsigdim}{\ysigdim}
\safemath{\rsig}{\bmr}
\safemath{\Adict}{\bA}
\safemath{\Adicttilde}{\widetilde{\Adict}}
\safemath{\Adictdim}{\outputdim\times\xsigdim}
\safemath{\avec}{\bma}
\safemath{\avectilde}{\tilde{\avec}}
\safemath{\Bdict}{\bB}
\safemath{\Bdicttilde}{\widetilde{\Bdict}}
\safemath{\Cdict}{\bC}
\safemath{\cvec}{\bmc}
\safemath{\Ddict}{\bD}
\safemath{\Ddictdim}{\ysigdim\times\xsigdim}
\safemath{\dvec}{\bmd}
\safemath{\Ddicttilde}{\widetilde{\bD}}
\safemath{\Bonb}{\bB}
\safemath{\bvec}{\bmb}
\safemath{\Bonbdim}{\ysigdim\times\ysigdim}
\safemath{\noise}{\bmn}
\safemath{\noisedim}{\ysigim}
\safemath{\err}{\bme}
\safemath{\errdim}{\ysigdim}
\safemath{\errset}{\setE}
\safemath{\nerr}{n_e}
\safemath{\delop}{\bP_\errset}
\safemath{\delopc}{\bP_{{\errset}^c}}

\safemath{\cplxi}{\imath}
\safemath{\cplxj}{\jmath}

\safemath{\dict}{\matD}
\safemath{\inputdim}{N}		
\safemath{\outputdim}{M}		
\safemath{\sparsity}{S}	
\safemath{\inputdimA}{{N_a}}	
\safemath{\inputdimB}{{N_b}}	
\safemath{\elemA}{{n_a}}	
\safemath{\elemB}{{n_b}}	
\safemath{\resA}{\matR_a}	
\safemath{\resB}{\matR_b}	
\safemath{\subD}{\matS} 
\safemath{\subA}{\matS_a} 
\safemath{\subB}{\matS_b} 
\safemath{\dicta}{\matA} 	
\safemath{\dictb}{\matB} 	
\safemath{\hollowS}{H}
\safemath{\hollowA}{H_a}
\safemath{\hollowB}{H_b}
\safemath{\cross}{Z}
\safemath{\coh}{\mu_d}			
\safemath{\coha}{\mu_a}			
\safemath{\cohb}{\mu_b}			
\safemath{\mubs}{\nu}	
\safemath{\cohm}{\mu_m} 
\safemath{\dictset}{\setD}	
\safemath{\dictsetp}{\dictset(\coh,\coha,\cohb)}	
\safemath{\dictsetgen}{\dictset_\text{gen}}
\safemath{\dictsetgenp}{\dictsetgen(\coh)}
\safemath{\dictsetonb}{\dictset_\text{onb}}
\safemath{\dictsetonbp}{\dictsetonb(\coh)}

\safemath{\leftside}{U}
\safemath{\rightsideA}{R_a}
\safemath{\rightsideB}{R_b}

\safemath{\indexS}{\setI_S} 

\safemath{\na}{n_a}			
\safemath{\nb}{n_b}			
\safemath{\coeffa}{p_i}	
\safemath{\coeffb}{q_j}	
\safemath{\seta}{\setP}		
\safemath{\setb}{\setQ}     
\safemath{\setw}{\setW}	
\safemath{\setz}{\setZ}	
\safemath{\cola}{\veca}		
\safemath{\colb}{\vecb}		
\safemath{\cold}{\vecd}		
\safemath{\inputvec}{\vecx} 	
\safemath{\error}{\vece}	
\safemath{\noiseout}{\vecz} 	
\safemath{\inputvecel}{x}
\safemath{\inputveca}{\vecx_a}
\safemath{\inputvecb}{\vecx_b}
\safemath{\outputvec}{\vecy}	
\safemath{\lambdamin}{\lambda_{\mathrm{min}}}

\safemath{\elltwo}{\ell_2}
\safemath{\ellone}{\ell_1}
\safemath{\ellzero}{\ell_0}
\safemath{\ellinf}{\ell_\infty}
\safemath{\ellinftilde}{\ell_{\widetilde\infty}}
\safemath{\licard}{Z(\coh,\coha,\cohb)}
\safemath{\xsol}{\hat{x}}
\safemath{\xbord}{x_b}		
\safemath{\xstat}{x_s}		
\safemath{\xstatLone}{\tilde{x}_s}
\safemath{\order}{\mathcal{O}} 
\safemath{\scales}{\Theta} 
\safemath{\ones}{\mathbf{1}} 
\safemath{\zeroes}{\mathbf{0}} 
\safemath{\thlone}{\kappa(\coh,\cohb)} 
\safemath{\constoneA}{\delta} 
\safemath{\constoneB}{\epsilon} 
\safemath{\nlarge}{L}				   
\safemath{\sumlarge}{S_\nlarge}
\safemath{\maxlarger}{P_\nlarge}	   
\safemath{\Pzero}{\textrm{P0}}	
\safemath{\Pone}{\textrm{P1}}
\safemath{\vecfir}{\vecw}			 
\safemath{\vecsec}{\vecz}
\safemath{\elvecfir}{w}              
\safemath{\elvecsec}{z}				 
\safemath{\nlargefir}{n}
\safemath{\normout}{\gamma}
\safemath{\auxfun}{h}
\safemath{\supp}{\textrm{supp}}

\safemath{\indexa}{\ell}
\safemath{\indexb}{r}
\safemath{\indexc}{i}
\safemath{\indexd}{j}

\safemath{\project}{P}

\usepackage{framed}

\renewcommand{\bSigma}{\boldsymbol{\Sigma}}
\newcommand{\bsigma}{\boldsymbol{\sigma}}
\newcommand{\Ie}{I^{\boldsymbol{\ast}}}
\newcommand{\hatIe}{\hat{I}^{\boldsymbol{\ast}}}

\newcommand{\secret}{\makebox[7pt][l]{\raisebox{-0.03cm}[0pt][0pt]{\twemoji[height=2.5mm,trim={0.1mm 0.1mm 0.1mm 0mm}, clip]{game_die}}}}
\newcommand{\largesecret}{\twemoji[height=3mm,trim={0.1mm 0.1mm 0.1mm 0mm}, clip]{game_die}}

\safemath{\bsfU}{\boldsymbol{\mathsf{U}}}
\safemath{\bsfV}{\boldsymbol{\mathsf{V}}}
\safemath{\bsfu}{\boldsymbol{\mathsf{u}}}
\safemath{\bsfv}{\boldsymbol{\mathsf{v}}}
\safemath{\sfsigma}{\mathsf{\sigma}}
\safemath{\bsfSigma}{\boldsymbol{\mathsf{\Sigma}}}

\safemath{\Cpar}{\bC_{\parallel}}
\safemath{\Corth}{\bC_{\orth}}
\safemath{\Ypar}{\bY_{\parallel}}
\safemath{\Yorth}{\bY_{\orth}}

\IEEEoverridecommandlockouts
\allowdisplaybreaks 

\newcommand*\tinygraycircled[1]{\Circled[inner color=white, fill color= gray, outer color=gray]{\footnotesize{\textnormal{#1}}}}
\newcommand*\scriptsizegraycircled[1]{\Circled[inner color=white, fill color= gray, outer color=gray]{\scriptsize{\textnormal{#1}}}}

\safemath{\Hj}{\bJ}
\safemath{\bsj}{\bmw}
\safemath{\sj}{w}
\safemath{\Ej}{E_w}
\safemath{\proxg}{\text{prox}_g}
\safemath{\rE}{\rho_{\textsf{\tiny{E}}}}
\safemath{\rP}{\rho_{\textsf{\tiny{P}}}}

\makeatletter
\newcommand\fs@spaceruled{\def\@fs@cfont{\bfseries}\let\@fs@capt\floatc@ruled
  \def\@fs@pre{\vspace{1mm}\hrule height.8pt depth0pt \kern2pt}%
  \def\@fs@post{\kern2pt\hrule\relax}%
  \def\@fs@mid{\kern2pt\hrule\kern2pt}%
  \let\@fs@iftopcapt\iftrue}
\makeatother

\begin{document}
\bstctlcite{IEEEexample:BSTcontrol} 

\title{Universal MIMO Jammer Mitigation}

\author{
\IEEEauthorblockN{Gian Marti and Christoph Studer}\\
\thanks{The authors are with the Department of Information Technology and Electrical Engineering, ETH Zurich. (email: marti@iis.ee.ethz.ch, studer@ethz.ch)}
\thanks{The work of GM and CS was supported in part by an ETH Research Grant.}
\thanks{A short version of this paper was presented at the 57th Asilomar Conference on Signals, Systems, and Computers
\cite{marti2023universal}.} 
\thanks{Emojis by Twitter, Inc.\ and other contributors are licensed under CC-BY~4.0.}
}
\maketitle
\thispagestyle{empty}

\begin{abstract}

Multi-antenna processing enables jammer mitigation through spatial filtering, provided
that the receiver knows the spatial signature of the jammer interference.
Estimating this signature is easy for barrage jammers that transmit continuously and with static signature, 
but difficult for more sophisticated jammers.
Smart jammers may deliberately suspend transmission
when the receiver tries to estimate their spatial signature, or they may use time-varying beamforming
to continuously change their spatial signature.
To deal with such smart jammers, we propose MASH, the first method that indiscriminately mitigates \emph{all} types of jammers.
Assume that the transmitter and receiver share a common secret. 
Based on this secret, the transmitter \emph{embeds} (with a time-domain transform) 
its signal in a secret subspace of a higher-dimensional space.
The receiver applies a reciprocal transform to the receive signal, 
which (i) \emph{raises} the legitimate transmit signal from its secret subspace 
and (ii) provably transforms \emph{any} jammer into a barrage 
jammer, making estimation and mitigation via multi-antenna processing straightforward. 
Focusing on the massive multi-user MIMO uplink, we present three MASH-based data detectors
and show their jammer-resilience via extensive simulations. 
We also introduce strategies for multi-user communication without a global secret
as well as methods that use computationally efficient embedding and raising transforms. 
 
\end{abstract}


\section{Introduction}
\IEEEPARstart{D}{espite} focusing on ultra-reliable communications as one of its main use-cases \cite{dahlman20205g}, 
5G remains critically vulnerable against 
jamming attacks \cite{lichtman20185g, girke2019towards, threatvectors2021cisa}. It is therefore imperative that next-generation wireless networks
are able to mitigate jammers.
Techniques such as direct-sequence spread spectrum (DSSS)~\cite{guanella1944dsss, madhow1994mmse}
or frequency-hopping spread spectrum (FHSS)~\mbox{\cite{tesla1903fhss, stark1985coding}} provide a certain degree of robustness against interference, but are no match for strong wideband jammers. 
Multi-antenna (MIMO) processing, in contrast, is able to completely remove jammer interference 
through spatial filtering~\cite{leost2012interference} and is therefore a promising path towards jammer-resilient communications.
However, MIMO jammer mitigation relies on accurate knowledge of the jammer's spatial 
signature (e.g., its subspace or its spatial covariance matrix) at the receiver. 
This signature is easy to estimate for barrage jammers (i.e., jammers that transmit continuously 
and with time-invariant signature):
The receiver can analyze the receive signal from a dedicated training period, 
which provides a representative snapshot of the jammer's spatial signature. 
A more sophisticated jammer, however, may thwart such a na\"ive approach, for example by deliberately suspending jamming
during the training period, by manipulating its spatial signature through time-varying beamforming, 
or by jamming only specific communication parts (e.g., control signals) and staying unnoticeable during other parts
\cite{ miller2010subverting,lichtman2016communications,lichtman20185g}. 
In all such cases, the signal received  during training does \emph{not} 
provide a representative snapshot of the jammer's spatial signature.

\subsection{Contributions}
We propose MASH (short for MitigAtion via Subspace Hiding), 
the first approach for MIMO jammer mitigation
that mitigates \emph{all} types of jammers, no 
matter how ``smart'' they are. 
MASH assumes that the transmitter(s) and the receiver share a common secret. 
Based on this secret, the transmitter(s) \emph{embed} their time-domain signals in a secret subspace of a higher-dimensional 
space using a linear time-domain transform; the receiver then applies an inverse linear time-domain transform to the
receive signal. We show that the receiver's transform
(i)~\emph{raises} the legitimate transmit signal from its secret subspace 
and (ii)~provably transforms \emph{any} jammer into a barrage jammer, which enables simplified mitigation.
To showcase the effectiveness of MASH, we consider data transmission in the massive multi-user MIMO uplink. 
We provide three MASH-based data detectors for this scenario, and we 
show their resilience against a range of jammers in extensive simulations. 

In addition to the material contained in the conference version \cite{marti2023universal}, 
this paper makes the following additional contributions:
We present a new result (\fref{prop:residual}) which proves that single-antenna barrage jammers can easily and effectively 
be mitigated. This result underpins our claim that, by transforming them into barrage jammers,
MASH makes \emph{all} jammers easy to mitigate. We also propose reciprocal MASH, a strategy
for using MASH in the practically relevant multi-user scenario in which the receiver shares 
a common secret with every transmitter, but where these secrets are pairwise distinct. 
Finally, we introduce computationally efficient transforms (which approximate the theoretically optimal ones) 
for embedding and raising.

\subsection{State of the Art}

The potential of MIMO processing for jammer mitigation is well known \cite{pirayesh2022jamming}. 
A variety of methods have been proposed that aim at mitigating barrage jammers 
\cite{marti2021hybrid, zhu2019mitigating, jiang2021efficient, chehimi2023machine, marti2021snips, yang2022estimation, he2022high, do18a, akhlaghpasand20a, akhlaghpasand20b, zeng2017enabling}, which either assume perfect knowledge of the jammer's spatial signature at the receiver 
\cite{marti2021hybrid, zhu2019mitigating, jiang2021efficient, chehimi2023machine} 
or exploit the jammer's stationarity by using a  training period 
\cite{marti2021snips, yang2022estimation, he2022high, do18a, akhlaghpasand20a, akhlaghpasand20b, zeng2017enabling}.
References\mbox{\cite{marti2021snips, yang2022estimation, he2022high}} estimate the jammer's spatial signature during a  dedicated
training period in which the legitimate transmitters are idle. 
References \cite{do18a, akhlaghpasand20a} estimate the jammer's spatial signature during a prolonged
pilot phase in which the jammer interference is separated from the legitimate transmit signals by projecting the receive
signals onto an unused pilot sequence that is orthogonal to the legitimate pilots. 
References~\cite{zeng2017enabling, akhlaghpasand20b} propose a jammer-resilient data detector directly as a function of certain spatial filters which are estimated during the pilot~phase. 
While all of these methods are only effective against barrage jammers, MASH transforms \emph{any}
jammer into a barrage jammer and therefore can be used to make the aforementioned methods 
effective also for the mitigation of non-barrage~jammers. 

It is widely known that jammers do not need to transmit in a time-invariant manner.
Several MIMO methods have been proposed for mitigating time-varying jammers
\cite{shen14a, yan2016jamming, hoang2021suppression, hoang2022multiple, marti2023maed, marti2023jmd}. 
References \cite{shen14a, yan2016jamming} consider a jammer that only jams when the legitimate
transmitters are active. Corresponding mitigation methods use a training period in which the legitimate
transmitters send pre-defined symbols that can be compensated for at the receiver. 
References\mbox{\cite{hoang2021suppression, hoang2022multiple}} assume a multi-antenna jammer that continuously
alters its spatial interference subspace through time-varying beamforming. 
Corresponding mitigation methods use a training period (in which the
legitimate transmitters are idle) recurrently and whenever the jammer subspace is believed to 
have changed substantially. 
References \cite{marti2023maed, marti2023jmd} consider general smart jammers. To mitigate such jammers, they  
propose a novel paradigm called joint jammer mitigation and data detection (JMD) in which the jammer subspace
is estimated and nulled jointly with detecting the transmit data over an entire communication frame. 
MASH is complementary to JMD, and the two can be productively combined (cf. \fref{sec:mash-maed}).

All of the above methods make stringent assumptions about the jammer's transmit behavior
(i.e., that the jammer is a barrage \cite{zeng2017enabling, jiang2021efficient, yang2022estimation, he2022high, chehimi2023machine, marti2021hybrid, marti2021snips, do18a, akhlaghpasand20a, akhlaghpasand20b} 
or reactive jammer \cite{shen14a, yan2016jamming}, or that the jammer's beamforming changes only slowly over time~\cite{hoang2021suppression, hoang2022multiple}), 
or they require solving a complex optimization problem while still being susceptible to certain types of jammers \cite{marti2023maed, marti2023jmd}.
Moreover, the number of jammer antennas is often required to be known in advance (see e.g., \cite{yan2016jamming, akhlaghpasand20a , marti2023jmd}).
MASH, in contrast, makes \emph{no} assumptions about the jammer's transmit behavior, 
does \emph{not} require solving an optimization problem, is effective against \emph{all} 
jammer types, and does \emph{not} require the number of jammer antennas to be known a priori.\footnote{It is important to not confound MASH with DSSS: DSSS transforms transmit symbols \emph{individually}, while MASH transforms entire blocks of symbols. 
Crucially, this means that DSSS does \emph{not} transform smart jammers into barrage jammers---this in stark contrast to MASH.}

\subsection{Notation}
Matrices and column vectors are represented by boldface uppercase and lowercase letters, respectively.
For a matrix~$\bA$, the transpose is $\tp{\bA}$, the conjugate transpose is $\herm{\bA}$, 
the pseudoinverse is $\pinv{\bA}$, 
the submatrix consisting of the columns (or rows) from $n$ through~$m$ is $\bA_{[n:m]}$ (or $\bA_{(n:m)}$, respectively),
the spectral norm is $\|\bA\|$, 
and the Frobenius norm is $\| \bA \|_\textnormal{F}$. 
For a vector $\bma$, the Euclidean norm is $\|\bma\|_2$.
The columnspace and rowspace of $\bA$ are $\text{col}(\bA)$ and $\text{row}(\bA)$, respectively.
Horizontal concatenation of two matrices $\bA$ and~$\bB$ is denoted by $[\bA,\bB]$; vertical concatenation is $[\bA;\bB]$.
The $k$th column and row of $\bA$ are $\bma_k$ and $\tp{\bma_{(k)}}$, respectively. 
The $N\!\times\!N$ identity matrix is $\bI_N$; 
the $M\!\times\!N$ all-zero matrix is $\mathbf{0}_{M\times N}$.
The compact singular value decomposition (SVD) of a rank-$R$ matrix $\bA\in\opC^{M\times N}$ is denoted
$\bA=\bsfU\bsfSigma\herm{\bsfV}$, where $\bsfU\in\opC^{M\times R}$ and $\bsfV\in\opC^{N\times R}$~have orthonormal 
columns and $\bsfSigma\in\opR^{R\times R}$ is a diagonal matrix whose diagonal entries are the positive singular 
values of $\bA$ in decreasing order. The main diagonal of $\bsfSigma$ is denoted by $\bsigma=\text{diag}(\bsfSigma)$. 
The distribution of a circularly-symmetric complex Gaussian random vector with covariance matrix $\bC$ is $\setC\setN(\mathbf{0},\bC)$.
$[1:N]$ are the integers from $1$ through $N$. 

\section{System Model}\label{sec:setting}

We consider jammer mitigation in communication systems with a multi-antenna receiver. 
Due to its relevance for next generation wireless networks, we focus on the multi-user (MU) MIMO uplink.
But we emphasize that our method can also be applied in single-user (SIMO) or point-to-point MIMO settings.
We consider a frequency-flat transmission model\footnote{An extension to frequency-selective channels 
with OFDM is possible \cite{marti2023single}.}
with the following input-output relation:
\begin{align}
	\bmy_k = \bH \bmx_k + \Hj \bsj_k + \bmn_k.  \label{eq:model}
\end{align}
Here, $\bmy_k\in\opC^B$ is the time-$k$ receive vector at the basestation (BS);
$\bH \in \opC^{B\times U}$ is the channel matrix of $U$ legitimate single-antenna user equipments (UEs) 
with time-$k$ transmit vector $\bmx_k\in\opC^U$;
$\bJ \in\opC^{B\times I}$ is the channel matrix of an \mbox{$I$-antenna} jammer
(or of multiple jammers, with $I$ being the total number of jammer antennas)
with time-$k$ transmit vector \mbox{$\bmw_k\in\opC^{I}$};
and $\bmn_k\sim\setC\setN(\mathbf{0},\No\bI_B)$ is i.i.d.\ circularly-symmetric complex white Gaussian  noise with 
per-entry variance~$\No$.
The channel matrices $\bH$ and $\bJ$ are assumed to stay constant for a coherence interval 
consisting of multiple channel uses. 
We assume that the jammer can vary its transmit characteristics.
For this, we use a model in which the jammer transmits 
\begin{align}
	\bsj_k = \bA_k \tilde\bsj_k
\end{align}
at time $k$, where, without loss of generality, 
$\Ex{}{\tilde\bmw_k\herm{\tilde\bmw_k}}=\bI_I$ for all $k$,
and where $\bA_k\in\opC^{I\times I}$ is a beamforming matrix that can change \emph{arbitrarily} as a function of~$k$. 
In particular,~$\bA_k$ can be the all-zero matrix (the jammer suspends jamming at time $k$), 
some of its rows can be zero (the jammer uses only a subset of its antennas at time~$k$), 
or it can be rank-deficient in another~way. 
There need not be any relation between~$\bA_k$ and $\bA_{k'}$ for~$k\neq k'$.
Our only restriction is that the number of jammer antennas $I$ is smaller than the number of BS antennas~$B$.

Unless noted otherwise, we assume that the UEs and the BS are jointly synchronized. 
Our proposed method operates on transmission frames of length $L$, 
where $L$ does not exceed the length $L_c$ of a coherence interval. 
Therefore, the channel matrices $\bH$ and~$\bJ$
are constant for the duration of a frame, 
and we restate~\eqref{eq:model} for a length-$L$ transmission frame as
\begin{align}
	\bY = \bH\bX + \bJ\bW + \bN, \label{eq:blockmodel}
\end{align}
where $\bY=[\bmy_1,\dots,\bmy_L]\in\opC^{B\times L}$ is the receive matrix, and
$\bX=[\bmx_1,\dots,\bmx_L]\in\opC^{U\times L}$~and $\bW = [\bA_1\tilde\bmw_1,\dots,\bA_L\tilde\bmw_L]\in\opC^{I\times L}$ are the corresponding 
transmit matrices. Since we do not impose any restriction on the matrices $\bA_k$, 
$\bW$ can be arbitrary and need in general not follow an explicit probabilistic model.

\section{On Barrage Jammers} \label{sec:barrage}
We prepare the ground for the exposition of MASH by considering a
class of jammers that is easy to mitigate: \emph{barrage jammers}.
Such jammers transmit noise-like stationary signals over the entire spectrum \cite{lichtman2016communications}, 
which appear as white (with respect to time) noise in the baseband representation \eqref{eq:blockmodel}.
Per~se, the notion of barrage jammers neither includes nor excludes the use of transmit beamforming. 

We now introduce a formal notion of barrage jammers. This notion explicitly 
allows for transmit beamforming,
as long as the beams stay constant for the duration of a communication~frame. 
Following our model in \fref{sec:setting}, we consider communication in frames of length $L$. 
During such a frame, a jammer transmits some matrix $\bW\in\opC^{I\times L}$ 
with corresponding receive interference $\bJ\bW \in \opC^{B\times L}$. 
We denote by $\Ie$ the rank of~$\bJ\bW$, i.e., $\Ie$ is the dimension of the interference space
which, depending on the jammer's transmit beamforming type, can be equal to or strictly smaller than $I$. 
We now consider the compact SVD of the interference, which we write as
\begin{align}
 	\bJ\bW = \bsfU\bsfSigma\herm{\bsfV}. \label{eq:barrage_svd}
\end{align}
We call $\bsfU\in\opC^{B\times \Ie}$ the \emph{spatial scope} of the jammer: 
The columns of this matrix form an orthonormal basis of the interference space in the spatial domain.   
We call ${\bsfV}\in\opC^{L\times\Ie}$ the \emph{temporal extension} of the jammer within the frame: 
its columns form an orthonormal basis of the interference space in the time domain. 
Finally, we call $\bsigma = \text{diag}(\bsfSigma)=\tp{[\sigma_1,\dots,\sigma_{\Ie}]}$ (with $\sigma_1\geq\dots\geq\sigma_{\Ie}>0$) the \emph{energy profile}, which 
determines how much energy the jammer allocates to the different dimensions in space and time. 
While the decomposition of the receive interference in \eqref{eq:barrage_svd} does not necessarily assume a probabilistic jamming model, 
our definition of barrage jammers is more narrow and explicitly requires a certain probabilistic behavior: \vspace{-1mm}

\begin{defi}[Barrage jammer] \label{def:barrage}
A barrage jammer is a jammer for which the columns of $\bsfV$ are distributed uniformly over the complex \mbox{$L$-dimensional} unit sphere.
\end{defi}
Barrage jammers are fully characterized by their spatial scope $\bsfU$ and their energy profile $\bsigma$.
Examples are a jammer that transmits $\bW$ with i.i.d. $\setC\setN(0,1)$ entries
or a jammer that transmits $\bW=\bma\tp{\bmw}$, with 
$\bma\in\opC^I$ and with $\bmw\sim\setC\setN(\mathbf{0},\bI_B)$.

Such a formal definition of barrage jammers allows us to see why 
barrage jammers can be mitigated easily and effectively: 
Consider a length-$L$ communication frame as in~\eqref{eq:blockmodel}, where the jammer is a barrage
jammer with compact SVD as given in~\eqref{eq:barrage_svd}. For simplicity, let 
us temporarily neglect the impact of thermal noise by assuming $\bN=\mathbf{0}$. 
If we insert a jammer training period of duration $R$ at the start of the frame, 
i.e., if the first $R$ columns of $\bX$ in \eqref{eq:blockmodel} are zero, then the corresponding receive signal is 
\begin{align}
	\bY_{[1:R]} = \bJ\bW_{[1:R]} = \bsfU\bsfSigma(\herm{\bsfV})_{[1:R]}. \label{eq:train}
\end{align}
Since the columns of $\bsfV$ are uniformly distributed over the complex $L$-dimensional unit sphere, 
the truncated temporal extension $(\herm{\bsfV})_{[1:R]}\in\opC^{\Ie\times R}$ 
has full rank $\min\{\Ie,R\}$ with probability one. 
So provided that $\Ie\leq R$, we have $\text{col}(\bY_{[1:R]})=\text{col}(\bsfU)$ with probability one. 
Thus, the receiver can find the jammer subspace $\text{col}(\bJ\bW)$ \emph{of the entire frame}
from the compact SVD $\bY_{[1:R]}\!=\!\tilde\bsfU\tilde\bsfSigma\herm{\tilde\bsfV}$ as 
$\text{col}(\bJ\bW)\!=\!\text{col}(\tilde\bsfU)$,
and it can mitigate the jammer \emph{for the entire frame} using a projection matrix 
$\bP=\bI_B - \tilde\bsfU\herm{\tilde\bsfU}$, which satisfies $\bP\bJ\bW=\mathbf{0}_{B\times L}$. 

A smart or dynamic jammer, in contrast, might stop jamming during these $R$ training samples, 
in which case \eqref{eq:train} amounts to \mbox{$\bY_{[1:R]} =\mathbf{0}_{B\times R}$} so that
the projection \mbox{$\bP=\bI_B$} is simply the identity, which will not mitigate
any subsequent interference of the jammer, since $\bP\bJ\bW_{[R+1:L]}=\bJ\bW_{[R+1:L]}$. 

Similarly, a dynamic multi-antenna jammer could at any given instant use only a subset 
of its antennas but~switch between different subsets over time, thereby changing the~spatial 
subspace of its interference. If $\text{col}(\bW_{[1:R]})\not\supseteq\text{col}(\bW_{[R+1:L]})$
and the columns of $\bJ$ are linearly independent, then 
$\text{col}(\bJ\bW_{[1:R]})\not\supseteq\text{col}(\bJ\bW_{[R+1:L]})$,
so the projection $\bP$ will not completely mitigate the jammer, 
$\bP\bJ\bW_{[R+1:L]}\neq\mathbf{0}_{B\times(L-R)}$.

\begin{figure}[tp]
\centering
\vspace{1mm}
\includegraphics[height=3.9cm]{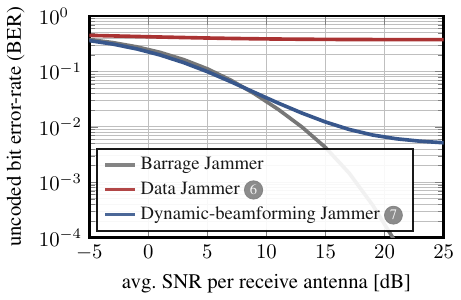}
\vspace{-2mm}
\caption{
Error-rate performance of a coherent communication receiver that~uses a training-period-based orthogonal projection 
for mitigating the jammer, least squares channel estimation, and LMMSE data detection.
The setup is described in \fref{sec:eval} and the $10$-antenna jammers are a barrage jammer transmitting 
i.i.d. $\setC\setN(0,1)$ symbols as well as the jammers $\scriptsizegraycircled{6}$ and $\scriptsizegraycircled{7}$
from \fref{sec:jammers}.}
\label{fig:different_jammers}
\vspace{-2mm}
\end{figure}

\begin{figure}[tp]
\centering
\includegraphics[height=3.8cm]{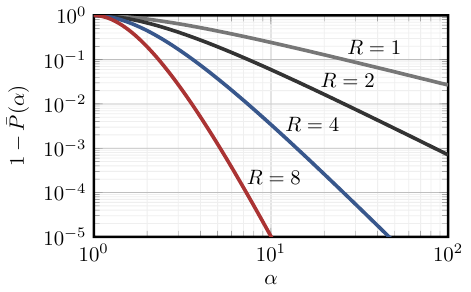}
\vspace{-1mm}
\caption{
Probability bound $1-\bar{P}(\alpha)$ that the residual interference power bound in \eqref{eq:intf_bound} does \emph{not} hold, 
as a function of the threshold $\alpha$, and for different redundancies $R$. 
As in \fref{sec:eval}, the frame length is $L=100$.
Note that the curves' asymptotic slope of descent (in doubly-logarithmic scale) coincides with $R$, 
so that the redundancy $R$ plays a role that is analogous to diversity.
}
\label{fig:bounds}
\vspace{-2mm}
\end{figure}

\fref{fig:different_jammers} shows the effectiveness of jammer mitigation with~an orthogonal projection based on a 
training period for a barrage jammer as well as for two non-barrage jammers 
(the simulations of \fref{fig:different_jammers} \emph{do} take into account the effects of thermal 
noise $\bN$).

For single-antenna jammers, we can also provide a theoretical guarantee for the mitigation of barrage jammers
in the presence of noise. Specifically, we have the following guarantee:\footnote{
We conjecture that an analogous result holds also for multi-antenna barrage jammers, 
but that case presents substantial additional analaytical difficulties.
}
\begin{prop} \label{prop:residual}
	Consider a single-antenna barrage jammer with channel vector $\bmj\in\opC^B$ 
	and frame transmit signal $\bmw^{\textnormal{T}}\in\opC^L$, so that the interference is 
	$\bmj\bmw^{\textnormal{T}}$. Let $\bN\in\opC^{B\times L}$ be the receive noise.
	Assume that (i) the first $R$ samples of the frame are used as a jammer training period during which the UEs are silent,
	(ii) the receiver estimates the jammer subspace as the leading left-singular vector $\bmu_1$ of the training period
	receive signal $(\bmj\bmw^{\textnormal{T}}+\bN)_{[1:R]}$, and (iii) the receiver mitigates the jammer in that frame
	with a projection $\bP=\bI_B-\bmu_1\bmu_1^{\textnormal{H}}$.
	Then, for any threshold $\alpha>1$, the residual interference power after projecting satisfies
	\begin{align}
		\frac{\|\bP\bmj\bmw^{\textnormal{T}}\|_\textnormal{F}^2}{L} \leq  12\,\alpha\, \frac{\|\bN_{[1:R]}\|_\textnormal{F}^2}{R} 
		\label{eq:intf_bound}
	\end{align}
	with probability at least $\bar{P}(\alpha)=1-\alpha^{-R} \left( \frac{L-R/\alpha}{L-R}\right)^{L-R}$.
\end{prop}

(Note that the guarantees of \fref{prop:residual} can be tightened at the expense of intuitiveness; 
the interested reader may consult the proof in \fref{app:proof1} for the details.)
For a single-antenna barrage jammer, the probability that the residual interference power $\|\bP\bmj\bmw^{\textnormal{T}}\|_\textnormal{F}^2/L$ after mitigation 
exceeds some value proportional to the training noise power $\|\bN_{[1:R]}\|_\textnormal{F}^2/R$ decreases polynomially 
with order $R$ (the length of the training period); cf. \fref{fig:bounds}.
Note that the bound for the residual jammer energy does not depend on the jammer transmit power, 
nor on the jammer's channel gain, since a larger jammer receive power yields a more accurate estimate of the 
jammer subspace, which leads to proportionally more effective~mitigation. 
\vspace{-2mm}

\section{MASH: Universal MIMO Jammer Mitigation\\ via Secret Temporal Subspace Embeddings} \label{sec:mash}
We are now ready to present MASH.
We assume that the UEs and the BS share a common secret~\secret.
Based on this secret, the UEs and the BS construct, in pseudo-random manner, 
a unitary matrix $\bC=f(\secret)$, which is uniformly (or \emph{Haar}) distributed over the set of all unitary $L\times L$
matrices.\footnote{See \cite[Sec.~1.2]{meckes2019random} on how to construct Haar distributed matrices.}
Since the jammer does not know the common secret \secret, it does not know $\bC$.
The matrix $\bC$ is then divided row-wise into two submatrices $\bC=[\Corth; \Cpar]$ 
with $\Corth\in\opC^{R\times L}$ and $\Cpar^{K \times L}$, 
where~$R$ and~$K$ are non-negative integers such that $R+K=L$. We call $R$ the \emph{redundancy}, and we require
that $R\geq \Ie$, i.e., that the redundancy is at least as big as the rank of the interference.\footnote{We do not require this dimension to be known a priori. The choice 
of $R$ simply limits number of interference dimensions that can be mitigated.
We note that the number of interference dimensions $\Ie$ that can be mitigated with spatial 
filtering is fundamentally (i.e., not just for MASH) limited by both the channel coherence time $L_c$ as well 
as the number of receive antennas $B$ (this is easy to show). The fact that MASH requires $\Ie\leq R \leq L_c$ is thus due to 
communication-theoretic limits, and not due to a drawback of MASH itself.}

The UEs use the matrix $\Cpar$ to \emph{embed} a \mbox{length-$K$} message signal $\bS\in\opC^{U\times K}$ 
in the secret $K$-dimensional subspace of~$\opC^L$ that is spanned by the rows of $\Cpar$ by transmitting
\begin{align}
	\bX = \bS\Cpar. \label{eq:embed}
\end{align}
The message signal $\bS$ could consist, e.g., of pilots for channel estimation, control signals, or data symbols. 
Embedding $\bS$ into the secret subspace requires no cooperation between the UEs: The $u$th UE simply transmits the 
$u$th row of $\bX$, which it can compute as $\tp{\bmx_{(u)}}=\tp{\bms_{(u)}}\Cpar$, 
where $\tp{\bms_{(u)}}$ is its own message signal. 
With~$\bX$ as in \eqref{eq:embed}, the receive matrix in \eqref{eq:blockmodel}~becomes 
\begin{align}
	\bY &= \bH \bX + \bJ\bW + \bN \label{eq:physical_io} \\
	&=\bH \bS \Cpar + \bJ\bW + \bN. \label{eq:mash_io}
\end{align}
Since $\bC$ is unknown to the jammer, $\bW$ is independent of~$\bC$. 
(This does not imply a probabilistic model for $\bW$---it only means that $\bW$ is no function of $\bC$.)
We make no other assump-tions about $\bW$. 
In particular, we also allow that $\bW$ might depend on $\bH$ and $\bJ$ (the jammer has full channel knowledge), 
or even on~$\bS$ (the jammer knows the message to be transmitted).

\begin{figure}[tp]
\centering
\vspace{1mm}
\includegraphics[width=0.95\columnwidth]{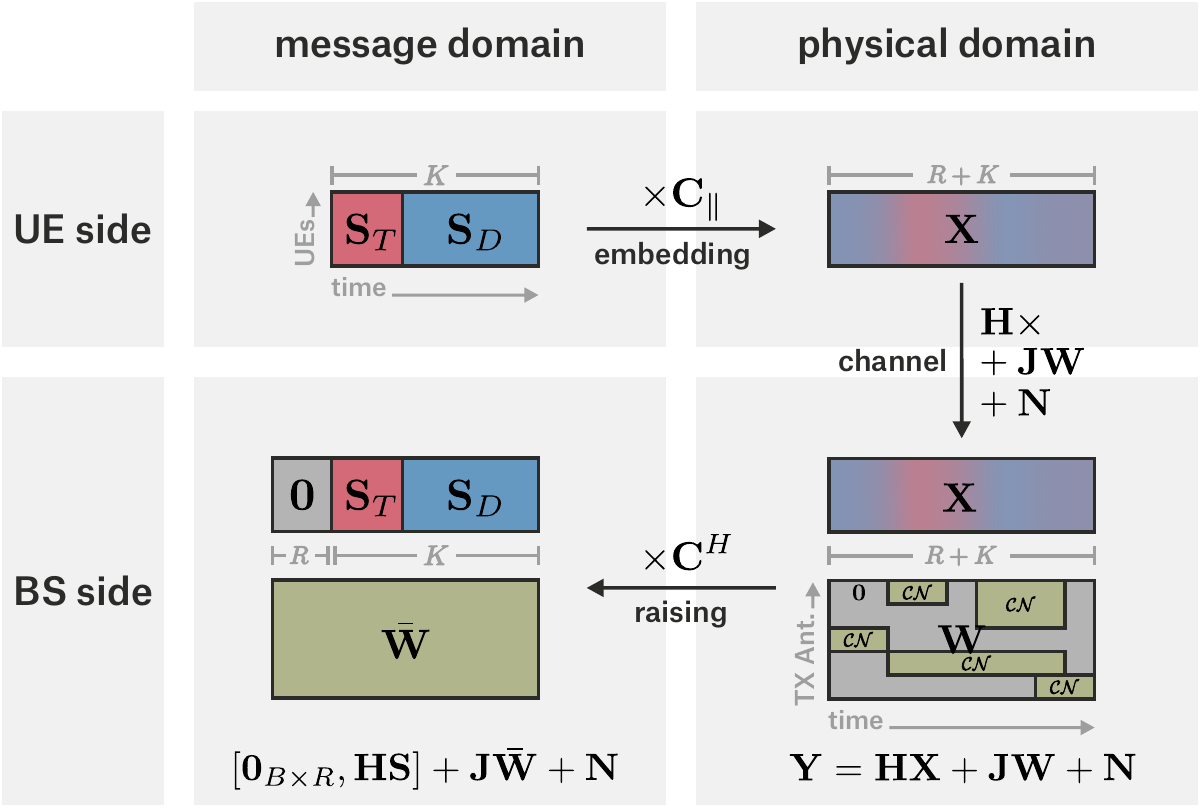}
\vspace{-2mm}
\caption{
Structural diagram of MASH. Here, the UE message signal $\bS$ consists of pilots $\bS_T$ (shown in red) 
and data symbols $\bS_D$ (shown in blue). 
The jammer is a multi-antenna jammer that dynamically switches different antennas off (shown in gray)
while transmitting Gaussian symbols using its active antennas (shown in green).
Raising transforms the jammer into a barrage jammer.
}
\label{fig:mash_diagram}
\vspace{-4mm}
\end{figure}

Having received $\bY$ as in \eqref{eq:mash_io}, 
the receiver \emph{raises} the signals by multiplying $\bY$ from the right
with $\herm\bC$, obtaining
\begin{align}
	\bar{\bY} &\triangleq \bY\herm{\bC}
	= \bH \bS \Cpar\herm{\bC} 
	 + \bJ\underbrace{\bW\herm{\bC}}_{~~\triangleq \bar{\bW}\!\!}
	 + \underbrace{\bN\herm{\bC}}_{~\,\triangleq \bar{\bN}\!\!} \\
	&= [\mathbf{0}_{B\times R}, \bH\bS] + \bJ\bar{\bW} + \bar{\bN}, \label{eq:raised}
\end{align}
where $\Cpar\herm{\bC}=[\mathbf{0}_{K\times R}, \bI_K]$ in \eqref{eq:raised} follows from the 
unitarity of $\bC=[\Corth; \Cpar]$. 
We obtain an input-output relation in which the channels $\bH$ and $\bJ$
are unchanged, the UEs are~idle during the first $R$ samples and transmit~$\bS$ in the remaining $K$ samples, 
the jammer transmits $\bar{\bW} = \bW\herm{\bC}$, and the noise  \mbox{$\bar{\bN} = \bN\herm{\bC}$} retains its statistics and
is i.i.d. circularly-symmetric complex Gaussian with per-entry variance~$\No$.
We call the domain of the input-output relation \eqref{eq:raised} the \emph{message domain}, 
which contrasts with the \emph{physical domain} of \eqref{eq:mash_io}; cf.~\fref{fig:mash_diagram}. 
We now state our key theoretical result; the proof is in \fref{app:app}: 
\begin{thm} \label{thm:barrage}
	Consider the decomposition of the physical domain jammer~interference $\bJ\bW$ in \eqref{eq:mash_io}
	into its spatial scope~$\bsfU$, its temporal extension $\bsfV$, and  its energy profile $\bsigma$. 
	Then the jammer interference $\bJ\bar{\bW}$ in the message domain \eqref{eq:raised} has identical spatial scope
	$\bar{\bsfU}=\bsfU$ and energy profile $\bar{\bsigma}=\bsigma$, 	but~its temporal extension $\bar{\bsfV}$ is a random 
	matrix whose columns are uniformly distributed over the complex $L$-dimensional unit sphere.
	Thus, $\bJ\bar{\bW}$ is barrage jammer interference according to~\fref{def:barrage}.
\end{thm}

Put simply, MASH transforms \emph{any} jammer into the (unique) barrage jammer 
with identical spatial scope and identical energy profile.\footnote{
The random behavior of barrage jammers (cf. \fref{def:barrage}) is induced by the randomness
of $\bC$ even if the jammer is deterministic in the physical~domain.
}
Furthermore, we see in the message-domain input-output relation \eqref{eq:raised} that 
MASH also conveniently deals with the legitimate transmit signals: 
The first~$R$ columns of the message domain receive signal $\bar{\bY}$ 
contain no UE signals and thus correspond 
to a jammer training period which can be used for 
constructing a jammer-mitigating filter that is effective for the entire frame (since the
message domain jammer is a barrage jammer).
The remaining $K$ columns of $\bar{\bY}$ correspond to the transmit signal $\bS$, which can be recovered 
using the jammer-mitigating filter from the training period. 

To understand how MASH can be used for MIMO jammer mitigation,
we consider the case of coherent data transmission,
for which we propose three different ways to proceed from~\fref{eq:raised}: 
(i)~orthogonal projection, (ii)~jammer-resilient LMMSE equalization, and (iii)~joint jammer mitigation and data detection. 
In coherent data transmission, the transmit signal $\bS=[\bS_T,\bS_D]$ consists of 
orthogonal pilots $\bS_T\in\opC^{U\times U}$ and data symbols $\bS_D\in\setS^{U\times (K-U)}$ taken from 
a constellation $\setS$.
To simplify the ensuing discussion, we define $[\bar{\bY}_J,\bar{\bY}_T,\bar{\bY}_D]\triangleq\bar{\bY}$ 
and rewrite \eqref{eq:raised} as three separate equations:
\vspace{-1mm}
\begin{align}
	\bar{\bY}_J &= \bJ\bar{\bW}_{J} + \bar{\bN}_J \in\opC^{B\times R} \label{eq:j_train}\\
	\bar{\bY}_T &= \bH\bS_T + \bJ\bar{\bW}_{T} + \bar{\bN}_T \in\opC^{B\times U} \label{eq:pilots}\\
	\bar{\bY}_D &= \bH\bS_D + \bJ\bar{\bW}_{D} + \bar{\bN}_D	 \in\opC^{B\times (K-U)}. \label{eq:data}
\end{align}
\vspace{-10mm}
\subsection{Orthogonal Projection} \label{sec:mash-pos}
The matrix $\bar{\bY}_J$ contains samples of a barrage jammer corrupted by white Gaussian noise $\bar{\bN}_J$, 
but not by any UE signals. We can thus use $\bar{\bY}_J$ to estimate the projection onto the orthogonal complement of the jammer's
spatial scope (cf.~\fref{sec:barrage}).
The maximum-likelihood estimate of the jammer's spatial scope is given by the $\Ie$ leading left-singular vectors
$[\bmu_1,\dots,\bmu_{\Ie}]\in\opC^{B\times\Ie}$ of $\bar{\bY}_J$ \cite{eckart1936approximation}.
The effective dimension $\Ie$ of the jammer interference is not known a priori.
This quantity can, however, be estimated directly from $\bar{\bY}_J$ as the number of singular values
that significantly exceed the threshold $\sqrt{B\No}$ which is expected due to thermal noise:\footnote{
We have $\Ex{}{\|\bN\|_\textnormal{F}^2}=BL\No$, and $\Ex{}{\|\bN\|_\textnormal{F}^2}$
equals the sum over $L$ squared nonzero singular values of $\bN$ (assuming $L\leq B$).
By dividing through $L$, the expected average of squared nonzero singular values is~$B\No$. 
}
\begin{align}
	\hatIe = \argmax_{\sigma_i(\bar{\bY}_J)>\beta\sqrt{B\No}} i, 
	\label{eq:est_Ie}
\end{align}
where $\sigma_i(\bar{\bY}_J)$ is the $i$th largest singular value of 
$\bar{\bY}_J$, and $\beta$ is a tuneable threshold parameter 
(for instance, $\beta=2$ is a sensible choice).
More complex methods, such as the Akaike information criterion (AIC) 
or the minimum description length (MDL) criterion, could also be used \cite{liavas2001behavior}.
With this estimate $\hatIe$ of $\Ie$, the estimate of jammer's
spatial scope is given by the $\hatIe$ leading left-singular vectors
$\bU(\bar\bY_J)\triangleq[\bmu_1,\dots,\bmu_{\hatIe}]\in\opC^{B\times\hatIe}$ of $\bar{\bY}_J$.
The corresponding projection matrix is therefore 
\begin{align}
	\hat\bP=\bI_B - \bU\herm{\bU}, \label{eq:mash_barrage_proj}
\end{align} 
where the dependence of $\bU=\bU(\bar\bY_J)$ on $\bar\bY_J$ is left implicit,
and the jammer can be mitigated in the pilot and data phases~via
\begin{align}
	\bar\bY_{\bP,T} &\triangleq \hat\bP\bar{\bY}_T = \underbrace{\hat\bP\bH}_{~\triangleq\bH_\bP\!\!\!\!\!\!\!}\bS_T 
	+ \hat\bP\bJ\bar{\bW}_{T} + \underbrace{\hat\bP\bar{\bN}_{T}}_{~~~\,\triangleq \bN_{\bP,T}\!\!\!\!\!\!} \\
	&\approx \bH_\bP\bS_T + \bN_{\bP,T}, \label{eq:pos_chest}
\end{align} 
and
\begin{align}
	\bar\bY_{\bP,D} &\triangleq \hat\bP\bar{\bY}_D = \hat\bP\bH\bS_D + \hat\bP\bJ\bar{\bW}_{D} + \hat\bP\bar{\bN}_{D} \\
	&\approx \bH_\bP\bS_D + \bN_{\bP,D}.	 \label{eq:pos_det}
\end{align}
The approximations in \eqref{eq:pos_chest} and \eqref{eq:pos_det} hold because
$\hat\bP\bJ\bar{\bW}_{T}\approx\mathbf{0}$ and $\hat\bP\bJ\bar{\bW}_{D}\approx\mathbf{0}$.
The BS thus obtains an input-output relation that is approximately jammer-free and consists of a virtual channel $\bH_\bP$
corrupted by~Gaussian noise with spatial distribution $\setC\setN(\mathbf{0},\No\bP)$.
The receiver can now estimate the virtual channel $\bH_\bP$, e.g., using least-squares (LS),~as follows:
\begin{align}
	\hat\bH_\bP = \bar\bY_{\bP,T}\pinv{\bS_T}.
\end{align}
The data symbols $\bS_D$ can then be detected using, e.g., the well-known 
linear minimum mean square error (LMMSE)~detector:
\begin{align}
	\hat\bS_D = \inv{( \herm{\hat{\bH}_\bP}\hat{\bH}_\bP + \No\bI_U )} \herm{\hat{\bH}_\bP} \bar\bY_{\bP,D}. \label{eq:p_lmmse}
\end{align}

\subsection{Jammer-Resilient LMMSE Equalization} \label{sec:mash-lmmse}
The orthogonal projection method from the previous subsection is highly effective. However, 
it requires the computation of a singular value decomposition (of $\bar{\bY}_J$) and the explicit 
estimation of the jammer interference dimension $\Ie$\!.
Both of these requirements can be avoided if, instead of an orthogonal projection, one directly uses
an LMMSE-type equalizer based on an estimate of the jammer's spatial covariance matrix
$\bC_{J}\triangleq\Ex{}{\bJ\bar{\bW}\herm{(\bJ\bar{\bW})}}$. 
Such an estimate can be obtained directly from $\bar{\bY}_J$ in \eqref{eq:j_train} as follows:
\begin{align}
	\hat\bC_J = \frac{1}{R} \bar{\bY}_J\herm{\bar{\bY}_J}. 
\end{align}
An ``LMMSE-type'' jammer-mitigating estimate  
of $\bH$ based on \eqref{eq:pilots} is then
(the impact of the thermal noise $\bar\bN_T$ is neglected) 
\begin{align}
	\hat\bH = \Big(\bI_B + \frac{1}{U} \hat\bC_J\Big)^{-1} \bar{\bY}_T\pinv{\bS_T}, \label{eq:mit_chest}
\end{align}
where the identity in the inverse represents the covariance matrix of the despreaded pilot signal,
and the LMMSE estimate of the data $\bS_D$ based on \eqref{eq:data} is
\begin{align}
	\hat\bS_D = \herm{\hat\bH}\big(\hat\bH\herm{\hat\bH}  + \No\bI_B + \hat\bC_J\big)^{-1} \bar{\bY}_D. \label{eq:mit_det}
\end{align}
This approach would avoid the calculation of an SVD, but would instead require us to invert two matrices of size $B\times B$ 
(in \eqref{eq:mit_chest} and in \eqref{eq:mit_det}).
However, using identities from \cite{petersen12a}, we can rewrite \eqref{eq:mit_chest} and \eqref{eq:mit_det} 
as
\begin{align}
	\hat\bH = \big(\bI_B - \bar{\bY}_J\inv{\big(UR\,\bI_R + \herm{\bar{\bY}_J}\bar{\bY}_J\big)}\herm{\bar{\bY}_J} \big)
	\bar{\bY}_T\pinv{\bS_T} \label{eq:small_chest}
\end{align}
and 
\begin{align}
	\! \bA &= \!\bigg( \!\No\bI_{U+R} + \begin{bmatrix} \herm{\hat\bH}\\ \frac{1}{\sqrt{R}}\herm{\bar{\bY}_J}\end{bmatrix}\!
		\begin{bmatrix} \hat\bH, & \!\!\!\!\!\frac{1}{\sqrt{R}}\bar{\bY}_J \end{bmatrix} \!\bigg)^{\!\!-1}\! 
		\begin{bmatrix} \herm{\hat\bH}\\ \frac{1}{\sqrt{R}}\herm{\bar{\bY}_J}\end{bmatrix} \label{eq:small_det}\! \\
	\! \hat\bS_D \!&= \bA_{(1:U)}\bar{\bY}_D.
\end{align}
So we only need to invert an $R\times R$ matrix (in \eqref{eq:small_chest}) and a $(U+R)\times(U+R)$ matrix
(in \eqref{eq:small_det}).
In contrast, the orthogonal projection approach would require computation of  an SVD for $B\times L$ matrix 
and the inverting a $U\times U$ matrix (in \eqref{eq:p_lmmse}). 
\vspace{-1mm}

\subsection{Joint Jammer Mitigation and Data Detection (JMD)} \label{sec:mash-maed}

JMD is a recent paradigm for smart jammer mitigation.
In JMD, the jammer subspace is estimated and nulled (through an orthogonal projection) \emph{jointly}
with detecting the transmit data over an entire transmission frame \cite{marti2023maed, marti2023jmd}. 
JMD operates by approximately solving a nonconvex optimization problem.
In principle, this paradigm can provide excellent performance \emph{even without} any jammer training period at all 
(corresponding to $R=0$). 
However, JMD suffers from several limitations:
(i) it performs poorly against jammers with very short duty cycle, 
(ii) it needs to know the dimension $\Ie$ of the jammer interference, and (iii)~it does not reliably converge to the correct solution if~$\Ie$ is large.
All of these shortcomings can either be completely avoided (in case of (i)) or substantially alleviated (in case of (ii) and (iii)) if JMD is combined with MASH:
(i)~MASH transforms jammers with short duty cycles into fully non-sparse barrage jammers, 
(ii)~in combination with non-zero redundancy ($R>0$), $\bar{\bY}_J$ can be used to estimate the dimension 
of the jammer interference, and
(iii)~$\bar{\bY}_J$ can be used for optimal initialization to improve performance against high-dimensional jammers.

We now show how MASH can be combined with the specific JMD-type methods SANDMAN and MAED from \cite{marti2023jmd, marti2024jmd}. 
SANDMAN uses a fixed LS channel estimate and then performs joint 
jammer mitigation and data detection by approximately solving an optimization 
problem that depends on the (unknown) transmit data and the (unknown) jammer-nulling projector. 
In contrast, MAED unifies not only jammer mitigation and data detection, but 
also channel estimation, by approximately solving an optimization problem 
that depends on the transmit data, the jammer-nulling projector, \emph{and} the UE channel matrix. 

The optimization problem to solve when combining MASH with SANDMAN is
\begin{align}
	\min_{\substack{
	\tilde\bP \,\in\, \mathscr{G}_{B-\hatIe}(\opC^B),\\
	\tilde\bS_D \,\in\, \setS^{U\times D}
	}} 
	\big\|\tilde\bP\big(\bar\bY_D - \hat\bH\tilde\bS_D\big)\big\|_\textnormal{F}^2.
	\label{eq:mash_s}
\end{align}
Here, $\mathscr{G}_{B-\hatIe}(\opC^B)=\{\bI_B-\bQ\pinv{\bQ}:\bQ\in\opC^{B\times\hatIe}\}$ is the Grassmannian manifold, 
i.e., the set of orthogonal projections onto $(B-\hatIe)$-dimensional subspaces of $\opC^B$, 
where $\Ie$ is estimated from $\bar{\bY}_J$ as in \eqref{eq:est_Ie}; 
and $\hat\bH=\bar{\bY}_T\pinv{\bS_T}$ is an LS channel estimate. 
This optimization problem can then be approximately solved as in \cite[Alg. 2]{marti2024jmd}. 
However, to improve robustness and speed of convergence even when facing many-antenna jammers, 
$\tilde\bP$ and $\tilde\bS_D$ should be initialized as in \eqref{eq:mash_barrage_proj} and \eqref{eq:p_lmmse}, 
respectively (and not to the initial values specified \mbox{in \cite[Alg.\,2]{marti2024jmd}}).
We call the algorithm that results from this procedure MASH-S; see \fref{alg:mash_s}. 
The step size $\tau^{(t)}$ on line~11 of \fref{alg:mash_s} can be chosen in a predetermined fashion or adaptively
(e.g., using the Barzilai-Borwein strategy~\cite{barzilai1988two}).
The function $\text{prox}_g$ on line 11 of \fref{alg:mash_s} is the so-called proximal operator \cite{goldstein16a}
of a suitably defined indicator function~$g$ (cf. \cite{marti2023jmd, marti2024jmd}) which acts entrywise on its input matrix. 
For QAM constellations, this entry-wise mapping can be written~as
\begin{align}
	\proxg(\tilde s; \tau) = 
	\begin{cases}
		\text{clip}\left(\frac{\tilde s}{1-\tau\alpha}; a\right) & \alpha\tau<1 \vspace{2mm}\\
		\displaystyle \arg\!\min_{\tilde x \in \setS } |\tilde s - \tilde x|^2 & \text{else,} 
	\end{cases}
	\label{eq:prox}
\end{align}
where $\text{clip}(z;a)$ clips the real and imaginary part of $z\in\opC$ to the interval $[-a,a]$, 
with $a$ being the ``height/width'' of the constellation $\setS$ 
(e.g., for unit symbol energy QPSK, $a=\sqrt2$).
The parameter $\alpha\geq0$ in \eqref{eq:prox} is a tunable parameter that pushes the symbol 
estimates towards the corners of the constellation, see \cite{marti2023jmd, marti2024jmd}.
In this paper, we choose $\alpha=2.5$. 

The optimization problem to solve when combining MASH with MAED is\vspace{-1mm}
\begin{align}
	\min_{\substack{
	\tilde\bP \,\in\, \mathscr{G}_{B-\hatIe}(\opC^B),\\
	\tilde\bH \,\in\, \opC^{B\times U},\\ 
	\tilde\bS_D \,\in\, \setS^{U\times D}
	}} 
	\big\|\tilde\bP\big([\bar\bY_T,\bar\bY_D] - \tilde\bH[\bS_T,\tilde\bS_D]\big)\big\|_\textnormal{F}^2,
\end{align}
where $\mathscr{G}_{B-\hatIe}(\opC^B)$ is as in \eqref{eq:mash_s}.
This problem can be transformed into an equivalent one that only depends on $\tilde\bP$ and~$\tilde\bS_D$ 
(see \cite[Sec.\,V-B]{marti2024jmd}) and then can be approximately solved as in \cite[Alg.\,3]{marti2024jmd}. 
As in the case of MASH-S, the robustness and speed of convergence can be improved if 
the algorithm is properly initialized using the information contained in $\bar{\bY}_T$, 
with the initial values of $\tilde\bP$ and $\tilde\bS_D$ being set using \eqref{eq:mash_barrage_proj} and~\eqref{eq:p_lmmse}, respectively. We call the algorithm that results from this procedure MASH-M; see \fref{alg:mash_m}.
The proximal operator $\text{prox}_g$ in line 11 of \fref{alg:mash_m} maps the first $T$ columns of its input 
matrix to $\bS_T$ and for the remaining columns acts identical to the proximal operator of \fref{alg:mash_s}.
\vspace{-5mm}

\floatstyle{spaceruled}
\restylefloat{algorithm}

\begin{algorithm}[tb]
  \caption{MASH-S}
  \label{alg:mash_s}
  \begin{algorithmic}[1]
	\setstretch{0.97}
    \Function{\textnormal{MASH-S}}{$\bar\bY_J, \bar\bY_T, \bar\bY_D, \bS_T, \beta\sqrt{B\No}, t_{\max}$}
    \State $\hat\bH = \bar\bY_T \pinv{\bS_T}$
    \State $\hatIe = \argmax_{\sigma_i(\bar{\bY}_J)>\beta\sqrt{B\No}} ~i$
    \State $\tilde\bP^{(0)}=\bI_B - \bU\herm{\bU}$ \hfill \textcolor{gray}{// cf. \eqref{eq:mash_barrage_proj}}
    \State $\tilde\bS_D^{(0)} = \inv{( \herm{\hat{\bH}}\tilde\bP^{(0)}\hat{\bH} + \No\bI_U )} \herm{\hat{\bH}} \tilde\bP^{(0)}\bar\bY_D$
    \For{$t=1$ {\bfseries to} $t_{\max}$}
		\State $\tilde\bE^{(t)} = [\bar\bY_T,\bar\bY_D] - \hat\bH [\bS_T,\tilde\bS_D^{(t-1)}]$
		\State $\tilde\bJ^{(t)} = \textsc{approxSVD}(\tilde\bE^{(t)}, \hatIe)$ 
		\hfill\textcolor{gray}{// cf. \cite[Alg.\,1]{marti2024jmd}} 
		\State $\tilde\bP^{(t)} = \bI_B - \tilde\bJ^{(t)}\pinv{(\tilde\bJ^{(t)})}$
		\State $\nabla f(\tilde\bS_D^{(t-1)}) = -2\,\herm{\hat\bH}\tilde\bP^{(t)}(\bar\bY_D - \hat\bH \tilde\bS_D^{(t-1)})$
		\State $\tilde\bS_D^{(t)} = \proxg\big(\tilde\bS_D^{(t-1)} - \tau^{(t)}\nabla f(\tilde\bS_D^{(t-1)}); \tau^{(t)}\big)$
    \EndFor
    \State \textbf{output:} $\tilde\bS_D^{(t_{\max})}$    
    \EndFunction    
  \end{algorithmic}
\end{algorithm}
\begin{algorithm}[tp]
  \caption{MASH-M}
  \label{alg:mash_m}
  \begin{algorithmic}[1]
	\setstretch{0.97}
    \Function{\textnormal{MASH-M}}{$\bar\bY_J, \bar\bY_T, \bar\bY_D, \bS_T, \beta\sqrt{B\No}, t_{\max}$}
    \State $\hat\bH = \bar\bY_T \pinv{\bS_T}$
    \State $\hatIe = \argmax_{\sigma_i(\bar{\bY}_J)>\beta\sqrt{B\No}} ~i$
    \State $\tilde\bP^{(0)}=\bI_B - \bU\herm{\bU}$ \hfill \textcolor{gray}{// cf. \eqref{eq:mash_barrage_proj}}
    \State $\tilde\bS^{(0)} = [\bS_T, \inv{( \herm{\hat{\bH}}\tilde\bP^{(0)}\hat{\bH} + \No\bI_U )} \herm{\hat{\bH}} \tilde\bP^{(0)}\bar\bY_D]$
    \For{$t=1$ {\bfseries to} $t_{\max}$}
		\State $\tilde\bE^{(t)} = [\bar\bY_T,\bar\bY_D](\bI_K - \tilde\bS^{(t-1)}{}^\dagger \tilde\bS^{(t-1)})$
    	\State $\tilde\bJ^{(t)} = \textsc{approxSVD}(\tilde\bE^{(t)}, \hatIe)$ 
		\hfill\textcolor{gray}{// cf. \cite[Alg.\,1]{marti2024jmd}} 
		\State $\tilde\bP^{(t)} = \bI_B - \tilde\bJ^{(t)}\pinv{(\tilde\bJ^{(t)})}$
		\State $\nabla f(\tilde\bS^{(t-1)}) = -2\herm{\big([\bar\bY_T, \bar\bY_D]\tilde\bS^{(t-1)}{}^\dagger\big)} \tilde\bP^{(t)}[\bar\bY_T, \bar\bY_D]$
		\par\hphantom{$\nabla f(\tilde\bS^{(t-1)}) =$}
		$\cdot(\bI_K - \tilde\bS^{(t-1)}{}^\dagger \tilde\bS^{(t-1)})$ 
		\State $\tilde\bS^{(t)} = \proxg\big(\tilde\bS^{(t-1)} - \tau^{(t)}\nabla f(\tilde\bS^{(t-1)}); \tau^{(t)}\big)$
    \EndFor
    \State \textbf{output:} $\tilde\bS^{(t_{\max})}_{[T+1:K]}$
    
    \EndFunction
    
  \end{algorithmic}
\end{algorithm}

\section{Practical Extensions and Modifications}\label{sec:practical}

\subsection{Individual Secrets: Reciprocal MASH} \label{sec:recip}
 
In the above description of MASH, all UEs were assumed to use the same secret~\secret.
From a practical perspective, this is a substantial drawback---the secret must be known by 
all UEs but the jammer must not know it.\footnote{As the saying goes: ``Two can keep a secret 
if one of them is dead.''}
We therefore propose \emph{reciprocal MASH}, a variant of MASH that can be
used with individual secrets, i.e., when every UE has its own secret known only to the BS and that UE.
In future cellular standards, such as 6G, such secrets could be distributed 
to the UEs, e.g., through the subscriber identity module (SIM card) \cite{jover2014enhancing}. 

The basic idea behind reciprocal MASH is simple: 
The BS detects the data for each UE separately, and while it does so, 
it treats the other UEs as additional jamming interference. 
This strategy has two disadvantages. First, it increases the computational 
complexity at the BS, since the BS now has to solve $U$ detection problems
(even if these detection problems are only single-user problems). 
Second, treating the other $U-1$ UEs as additonal interferers is suboptimal 
for detection and requires that the redundancy satisfies $R\geq \Ie+U-1$ 
rather than just $R\geq\Ie$. 
However, these disadvantages contrast with two significant advantages. 
The first advantage is the already discussed leap in practicality
with regards to how the required secrets can be distributed to the UEs.
The second advantage is the following: Since the BS regards all other 
UEs as additional interferers when detecting the data of any given UE, 
the UEs do not need to be 
synchronized to a joint schedule.
This is advantageous for the following reason: 
To enable the joint synchronization of multiple UEs in existing cellular systems, 
synchronization signals have to be sent by the BS \cite{omri2019synchronization}.
UEs can therefore be synchronized jointly---even for the uplink---only if they can 
correctly detect the synchronization signals in the downlink.\footnote{
Jammer mitigation in the downlink is an even more formidable problem 
than in the uplink, because UEs may not have more antennas
than a jammer. However, in many scenarios (e.g., internet of things or sensor networks), 
it may already be beneficial if a unidirectional 
communication link can be established.  
}
However, to the best of our knowledge, the only synchronization method that is resilient against
universal jamming attacks is \cite{marti2024jass}, which is a method
for uplink synchronization, and which thus cannot synchronize 
multiple UEs jointly. Reciprocal MASH therefore enables multi-user MIMO even under the limitations
of currently existing methods for jammer-resilient synchronization.

Reciprocal MASH works as follows: 
Let $\tp{\bms_{(u)}}\in\opC^{1\times K}$ and $\bC_u = [\Corth{_u};\Cpar{_u}]\in\opC^{L\times L}$ be the transmit signal and the embedding matrix for the $u$th UE.
Then the $u$th UE transmits 
\begin{align}
	\tp{\bmx_{(u)}}=\tp{\bms_{(u)}}\Cpar{_u}
\end{align}
and the BS receives\footnote{The receive signal $\bY_u$ depends on the index $u$ of the considered UE because
the BS is synchronized to the frame alignment of the $u$th UE, which may differ
from the frame alignment of the other UEs. If all UEs are synchronized to the same frame alignment, then 
$\bY_u$ does not depend~on~$u$.}
\begin{align}
	\bY_u = \bmh_u \tp{\bmx_{(u)}} + \sum_{\substack{u'=1\\u'\neq u}}^U
	 \bmh_{u'} \tp{\breve\bmx_{(u')}} + \bJ\bW_u + \bN_u. \label{eq:rec_io1}
\end{align}
Here, $\tp{\breve\bmx_{(u')}}\in\opC^{1\times L}$ is the signal that the 
$u'$th UE transmits while the $u$th UE transmits $\tp{\bmx_{(u)}}$. 
However, because we do not require that the UEs are jointly
synchronized, it is generally not the case that 
$\tp{\breve\bmx_{(u')}}=\tp{\bms_{(u')}}\Cpar{_{u'}}$. Instead, 
$\tp{\breve\bmx_{(u')}}$ can overlap two of the $u'$th UE's transmit frames
and thus may be any contiguous length-$L$ subvector of 
$[\tp{\bms_{(u'),1}}\Cpar{_{u'}},\tp{\bms_{(u'),2}}\Cpar{_{u'}}]$,
where $\tp{\bms_{(u'),1}}$ and $\tp{\bms_{(u'),2}}$ are the $u'$th UE's
messages which overlap with the current frame of the $u$th~UE. 
By defining $\breve\bX_{-u}\in\opC^{(U-1)\times L}$ as the matrix whose rows consist of the row vectors
$\tp{\breve\bmx_{(u')}}$ for all $u'\in[1:U]\setminus\{u\}$, 
and $\breve\bH_{-u}\in\opC^{B\times(U-1)}$ as the matrix whose columns consist of the channel vectors 
$\bmh_{u'}$ for all $u'\in[1:U]\setminus\{u\}$,
we can rewrite \eqref{eq:rec_io1} as
\begin{align}
	\bY_u &= \bmh_u \tp{\bmx_{(u)}} +
	\underbrace{\begin{bmatrix}
		\breve\bH_{-u} & \bJ
	\end{bmatrix}}_{\triangleq \breve\bJ_u\in\opC^{B\times (I+U-1)}\hspace{-21mm}}
	\qquad\cdot\qquad
	\underbrace{\begin{bmatrix}
		\breve\bX_{-u} \\ \bW_u
	\end{bmatrix}}_{\triangleq \breve\bW_u\in\opC^{(I+U-1)\times L}\hspace{-22mm}}
	+ \bN_u \\
	&= \bmh_u \tp{\bmx_{(u)}} + \breve\bJ_u \breve\bW_u + \bN_u. \label{eq:rec_io2}
\end{align}
The BS can now raise the $u$th UE's message by multiplying $\bY_u$
with $\herm{\bC_{u}}$, obtaining
\begin{align}
	\bar\bY_u &\triangleq \bY_u \herm{\bC_{u}} 
	= \bmh_u \tp{\bms_{(u)}}\Cpar{_u}\herm{\bC_{u}}
	+ \breve\bJ_u \underbrace{\breve\bW_u \herm{\bC_{u}}}_{\triangleq \bar\bW_u }
	+ \underbrace{\bN\herm{\bC_{u}}}_{\triangleq \bar\bN_u} \\
	&= \bmh_u[\mathbf{0}_{1\times R}, \tp{\bms_{(u)}}] + \breve\bJ_u \bar\bW_u + \bar\bN_u,
\end{align}
in direct analogy to \eqref{eq:raised}. The raised receive signal for the $u$th UE
corresponds thus to a single-user signal (which includes a length-$R$ jammer training period) 
that is contaminated by a ($I+U-1$)-antenna barrage ``jamming'' signal.\footnote{
The effective rank of the interference $\breve\bJ_u \bar\bW_u$ may be less than $I+U-1$
in the same way that $\Ie$ may be smaller than $I$, see \fref{sec:barrage}.}
The message of the $u$th UE can be detected using the methods of \fref{sec:mash}, 
and the entire process is repeated for every $u\in[1:U]$.

\subsection{Efficient Embedding/Raising Transforms} \label{sec:efficient}
For \fref{thm:barrage} in \fref{sec:mash}, we have assumed that the embedding matrix $\bC=f(\secret)$ 
is Haar distributed, i.e., uniformly distributed over the set of unitary $L\times L$ matrices. 
A Haar distributed matrix can be obtained as follows: Draw a matrix with i.i.d. $\setC\setN(0,1)$ entries 
and compute its QR-decomposition using the Gram-Schmidt procedure. The resulting $\bQ$-matrix 
is Haar distributed \cite[Sec.~1.2]{meckes2019random}. 
Unfortunately, this approach is not practical as it exhibits high computational complexity. 
If $f$ is a deterministic function that converts the secret~$\secret$
first into a Gaussian matrix and from there into a Haar matrix, then $\secret$ needs to contain 
an infinite amount of entropy. 
Furthermore, to prevent the jammer from learning the matrix $\bC$ over time, this Haar distributed transform 
matrix should be replaced frequently (ideally after every~use). But computing the QR-decomposition has a complexity of 
$O(L^3)$, so the computational cost of doing so would be prohibitive. 

To address these two issues, we propose to use transform matrices for embedding and raising
that are approximately Haar distributed. Specifically, we approximate a Haar matrix~as
\begin{align}
	\bC = \bF \bD_1 \bF \bD_2 \bF, \label{eq:approx_C}
\end{align}
where $\bD_1$ and $\bD_2$ are diagonal random matrices that have 
i.i.d. $\text{Unif}\{-1,+1\}$ diagonal entries (i.e., diagonal matrices whose diagonals are Rademacher distributed), 
and where $\bF$ is either the discrete fourier transform (DFT) matrix or the Walsh-Hadamard
matrix. Such matrices approximate Haar matrices in the sense that, 
like Haar matrices, they are restricted isometry property (RIP) optimal 
in certain regimes \cite{ailon2014fast}. Concatenations of 
Walsh-Hadamard and random Rademacher diagonal matrices have also been 
used to approximate random rotations\footnote{
The random ``rotations'' in \cite{andoni2015practical} are random Gaussian 
matrices that are only approximately unitary, but that behave similarly to Haar matrices\mbox{\cite{borel1914introduction, sutton2005stochastic}.}
}
in \cite{andoni2015practical}. 
Note that using a matrix $\bC$ as in \eqref{eq:approx_C} for embedding and raising does
not even require $\bC$ to be computed explicitly. Instead, 
multiplying with $\Cpar$ (embedding) and with $\herm{\bC}$ (raising) can be
implemented by a series of fast Fourier/Hadamard transforms that alternates with appropriate element-wise sign changes.
Using fast transforms also reduces the computational complexity 
of raising/embedding itself; see \fref{tbl:complexity}.

\begin{table}[tp]
\centering
\setstretch{1.1}
\caption{Computational Complexity of Embedding and Raising}
\small
\label{tbl:complexity}
\begin{tabular}{lcc} 
\toprule
& Exact Haar\!  & Approximate Haar\\
\midrule
Construction of $\bC^{\footnotesize a}$	 	& $O(L^3)$ & 0 \\ 
Embedding (per UE)	& $O(LK)$ & $O(L\log L)$ \\ 
Raising 		& $O(BL^2)$	& $O(BL\log L)$ \\
\bottomrule
\multicolumn{3}{l}{\parbox[t]{0.89\columnwidth}{
\footnotesize
$^{\footnotesize a}$The impact on complexity of constructing a pseudorandom Gaussian matrix or a
Rademacher vector is neglected.}
}
\end{tabular}	
\end{table}

%

\section{Experimental Evaluation} \label{sec:eval}

\subsection{Simulation Setup} \label{sec:setup}

We simulate the uplink of a massive multi-user (MU) MIMO system with $B=64$ antennas at the BS and $U=16$ UEs. 
The channel vectors are generated using QuaDRiGa \cite{jaeckel2014quadriga} with the 
3GPP 38.901 urban macrocellular (UMa) channel model~\cite{3gpp22a}. 
The carrier frequency is $2$\,GHz and the BS antennas are arranged as a uniform linear array (ULA)
with half-wavelength spacing. The UEs and the jammer are distributed randomly at distances 
from $10$\,m to $250$\,m in a $120^\circ$ angular sector in front of the BS, with a minimum angular 
separation of $1^\circ$ between any two UEs as well as between the jammer and any UE. 
The jammer can be a single- or a multi-antenna jammer (see \fref{sec:jammers}).
The antennas of multi-antenna jammers are arranged as a half-wavelength ULA that is 
facing the BS's direction. All antennas are assumed to be omnidirectional and the UEs are assumed to use $\pm3$\,dB power~control. 

As in \fref{sec:mash}, we consider coherent data transmission. 
The framelength is $L=100$, and is divided into a redundancy of $R=16$, 
$T=16$ pilot samples, and $L-R-T=68$ data samples. 
The transmit constellation $\setS$ is QPSK, and the pilots $\bS_T$ 
are chosen as a $16\times16$ Hadamard matrix (normalized~to unit symbol energy).
We characterize the strength of the jammer interference relative to the 
strength of the average UE via
\begin{align} 
	\rho \define \frac{\|\bJ\bW\|_\textnormal{F}^2}{\frac1U\Ex{\bS}{\|\bH\bX\|_\textnormal{F}^2}}.
\end{align}
The average signal-to-noise ratio (SNR) is defined as
\begin{align}
\textit{SNR} \define \frac{\Ex{\bS}{\|\bH\bX\|_\textnormal{F}^2}}{\Ex{\bN}{\|\bN\|_\textnormal{F}^2}}.
\end{align}
As performance metrics, we use uncoded bit error rate (BER) as well as 
the modulation error ratio (MER)
\begin{align} 
	\textit{MER}\triangleq \frac{\mathbb{E}\big[\|\hat\bS_D - \bS_D\|_\textnormal{F}\big]}{\mathbb{E}\big[\|\bS_D\|_\textnormal{F}\big]},
\end{align}
which is a surrogate for error vector magnitude (EVM) \cite{3gpp21a}. 
Recall that the 3GPP 5G NR technical specification requires 
an EVM below 17.5\%\cite[Tbl. 6.5.2.2-1]{3gpp21a} for QPSK transmission.

\subsection{Methods and Baselines} \label{sec:methods}
\begin{figure}[tp]
\centering
\subfigure[MASH]{\includegraphics[height=1.05cm]{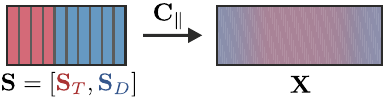}\label{fig:mash}}
\!\hfill\!
\subfigure[Baselines]{\includegraphics[height=1.05cm]{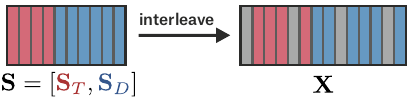}\label{fig:baselines}}
\caption{MASH (left) maps a length-$K$ signal $\bS=[\bS_T,\bS_D]$ to a length-$L$ signal $\bX$ by multiplying
with $\Cpar$. The baselines (right) map a length-$K$ signal $\bS=[\bS_T,\bS_D]$ to a length-$L$ signal 
by interleaving it with evenly spread zero-symbols (shown in gray) that serve as jammer training period.~For 
the baselines, we denote the receive samples corresponding to the jammer training period, 
pilot phase, and data phase as $\bY_J$, $\bY_T$ and~$\bY_D$, respectively. 
These matrices are the non-MASH counterparts of $\bar\bY_J,\bar\bY_T$ and $\bar\bY_D$ 
in \eqref{eq:j_train}-\eqref{eq:data}.}
\label{fig:mash_vs_baselines}
\end{figure}

We compare MASH against baselines which, instead of embedding the pilots
and data symbols in a secret higher-dimensional space, transmit them in the conventional
way, but interleave them with $R$ zero-symbols that are evenly distributed over the frame
and serve as jammer training period (cf. \fref{fig:mash_vs_baselines}).

\subsubsection*{\textbf{MASH-L}}
This receiver operates as described in Sec. \ref{sec:mash-lmmse}.

\subsubsection*{\textbf{MASH-M}}
This receiver operates as described in Sec. \ref{sec:mash-maed} and uses $t_{\max}=10$ algorithm iterations.

\subsubsection*{\textbf{JL}}
This baseline operates on a jammerless (JL) system. The transmitter operates as
depicted in \fref{fig:baselines}. The receiver ignores
the receive samples of the jammer training period and performs LS 
channel estimation and LMMSE data detection. 

\subsubsection*{\textbf{Unmitigated (Unm.)}}
This baseline does not mitigate the jammer. The transmitter operates as
depicted in \fref{fig:baselines}. The receiver ignores
the receive samples of the jammer training period and performs LS 
channel estimation and LMMSE data detection as one would in a jammerless environment.

\subsubsection*{\textbf{LMMSE}}
This baseline is identical to MASH-L, except that it does not use secret subspace embeddings. 
The transmitter operates as depicted in \fref{fig:baselines}.
The receiver estimates the jammer's spatial covariance matrix
using the receive samples $\bY_J$ from the training period, 
$\hat\bC_J = \frac1R \bY_J\herm{\bY_J}$, 
and then performs jammer-mitigating channel estimation and data detection analogous to 
\eqref{eq:mit_chest} and \eqref{eq:mit_det}. 

\subsubsection*{\textbf{MAED}}
This baseline is identical to MASH-M, except that it does not use secret subspace embeddings. 
The transmitter operates as depicted in \fref{fig:baselines}, and 
the receiver operates as in \fref{alg:mash_m}, except that its inputs are $\bY_J, \bY_T, \bY_D$ 
instead of $\bar\bY_J, \bar\bY_T, \bar\bY_D$. It uses $t_{\max}=10$ algorithm iterations as well.

For the sake of figure readability, we omit the orthogonal projection variant of \fref{sec:mash-pos}
from our experiments, which performs similarly as MASH-L. 
None of the methods is given a~priori knowledge about the dimension $\Ie$ of the interference space. 
The methods that require such knowledge estimate $\Ie$ as the number of the singular values of 
$\bar\bY_J$~(\mbox{MASH-M}) or of~$\bY_J$ (MAED) that exceed $2\sqrt{B\No}$; cf.~\eqref{eq:est_Ie}.

\begin{figure*}[tp]
\centering
\!\!\!\!\!\!
\subfigure[single-antenna barrage jammer \tinygraycircled{1}]{
\includegraphics[height=3.35cm]{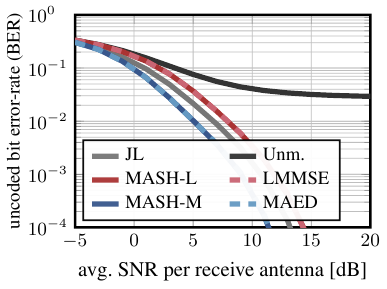}
\label{fig:barrage}
}\!\!\!
\subfigure[single-antenna data jammer \tinygraycircled{2}]{
\includegraphics[height=3.35cm]{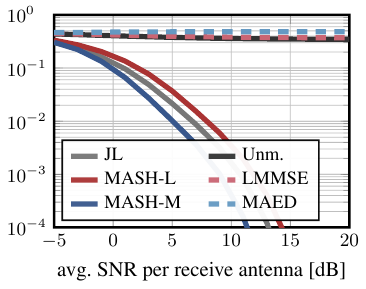}
\label{fig:data}
}\!\!\!
\subfigure[single-antenna pilot jammer \tinygraycircled{3}]{
\includegraphics[height=3.35cm]{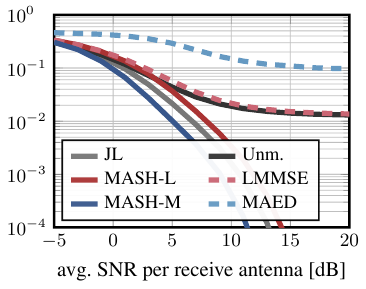}
\label{fig:pilot}
}\!\!\!
\subfigure[single-antenna sparse jammer \tinygraycircled{4}]{
\includegraphics[height=3.35cm]{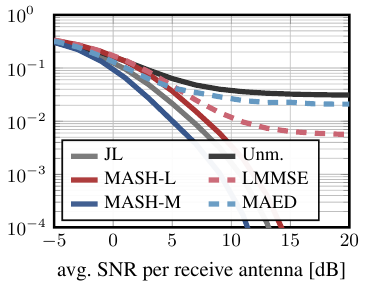}
\label{fig:sparse}
}\!\!\!\!
\!\!\!\!\!\!
\subfigure[multi-ant. eigenbeamf. jammer \tinygraycircled{5}]{
\includegraphics[height=3.35cm]{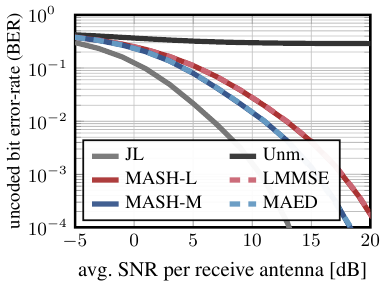}
\label{fig:eigenbeamforming}
}\!\!\!
\subfigure[multi-antenna data jammer \tinygraycircled{6}]{
\includegraphics[height=3.35cm]{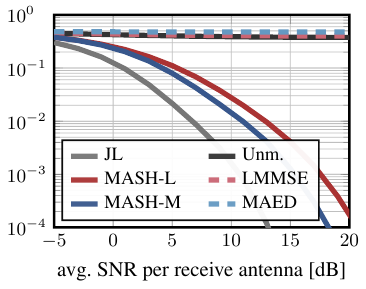}
\label{fig:multi-data}
}\!\!\!
\subfigure[multi-ant. dynamic jammer \tinygraycircled{7}]{
\includegraphics[height=3.35cm]{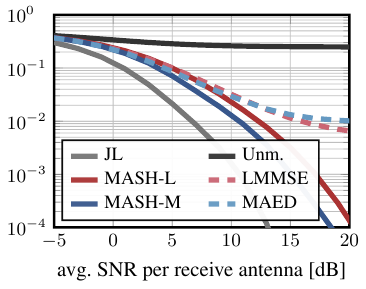}
\label{fig:dynamic}
}\!\!\!
\subfigure[multi-antenna repeat jammer \tinygraycircled{8}]{
\includegraphics[height=3.35cm]{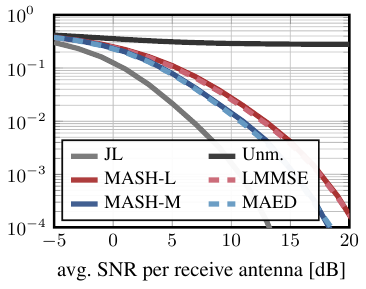}
\label{fig:repeater}
}\!\!\!\!
\caption{
Bit error rate (BER) performance of the different MASH receivers and baselines for eight different types of jammers. 
}
\vspace{-2mm}
\label{fig:jammers}
\end{figure*}

\subsection{Jammer Types} \label{sec:jammers}
The transmit signals of all jammers are normalized to \mbox{$\rho=30\,$dB.}
All multi-antenna jammers have $I=10$ antennas. 
\subsubsection*{\tinygraycircled{1} Single-antenna barrage jammer} 
This jammer transmits i.i.d. $\setC\setN(0,1)$ symbols for all samples.
\subsubsection*{\tinygraycircled{2} Single-antenna data jammer}
This jammer transmits i.i.d. $\setC\setN(0,1)$ symbols 
in all samples in which the UEs transmit data symbols\footnote{This refers to the samples of 
the baseline methods as in \fref{fig:baselines}. The subspace embeddings of MASH entail that 
data symbols are not transmitted during specific samples; cf. \fref{fig:mash}.
The same applies to jammers \tinygraycircled{3} and~\tinygraycircled{6}.}
and is idle for all other samples. 
\subsubsection*{\tinygraycircled{3} Single-antenna pilot jammer}
This jammer transmits i.i.d. $\setC\setN(0,1)$ symbols 
during the UE pilot transmission period and is idle for all other samples.
\subsubsection*{\tinygraycircled{4} Single-antenna sparse jammer}
This jammer transmits i.i.d. $\setC\setN(0,1)$ symbols 
in a fraction of $\alpha=0.1$ of \mbox{non-contigu}-ous randomly selected samples and is idle for all other~samples.
\subsubsection*{\tinygraycircled{5} Multi-antenna eigenbeamforming jammer}
This jammer is assumed to have full knowledge of $\bJ=\bsfU\bsfSigma\herm{\bsfV}$ and 
uses eigenbeamforming to transmit $\bW=\bsfV\tilde\bW$, where the entries of $\tilde\bW$
are i.i.d. $\setC\setN(0,1)$. According to Definition~\ref{def:barrage}, this jammer is
a barrage jammer.
\subsubsection*{\tinygraycircled{6} Multi-antenna data jammer}
This jammer transmits i.i.d. random vectors $\bmw_k\sim\setC\setN(\mathbf{0},\bI_I)$ for all samples $k$ in which 
the UEs transmit data symbols and is idle in all other~samples. 
\subsubsection*{\tinygraycircled{7} Multi-antenna dynamic-beamforming jammer}
At any given instance $k$, this jammer uses only a subset of its antennas, but it 
uses dynamic beamforming to change its antennas over time. 
Specifically, its time-$k$ beamforming matrix $\bA_k$ contains at most eight non-zero rows 
(the index set of these rows is generated uniform at random) whose entries are drawn i.i.d.~at random from $\setC\setN(0,1)$.
The matrix $\bA_{k+1}$ is equal to $\bA_k$ with probability $0.95$, and with probability
$0.05$, it is redrawn at random. The vectors $\tilde\bmw_k$ are drawn from 
$\setC\setN(\mathbf{0},\bI_I)$ for all~$k$. 
\subsubsection*{\tinygraycircled{8} Multi-antenna repeat jammer}
This jammer is assumed to have sensing capabilities that allow it to perfectly detect the UE transmit signal. 
The jammer then simply repeats the transmit signal $\bX_{(1:i)}$ (cf. \eqref{eq:embed}) of the first $I$ UEs with a delay of 
one sample, i.e., $\bW = [\mathbf{0}_{I\times 1},\bX_{(1:I),[1:L-1]}]$.

\begin{figure*}
\centering
\!\!\!\!
\subfigure[single-antenna barrage jammer \tinygraycircled{1}]{
\includegraphics[height=3.35cm]{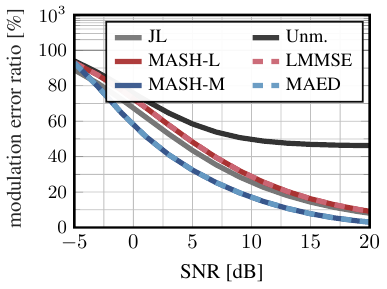}
}
\subfigure[single-antenna data jammer \tinygraycircled{2}]{
\includegraphics[height=3.35cm]{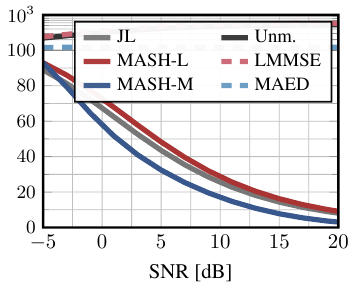}
}
\subfigure[single-antenna pilot jammer \tinygraycircled{3}]{
\includegraphics[height=3.35cm]{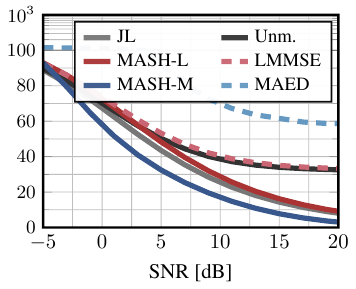}
}
\subfigure[single-antenna sparse jammer \tinygraycircled{4}]{
\includegraphics[height=3.35cm]{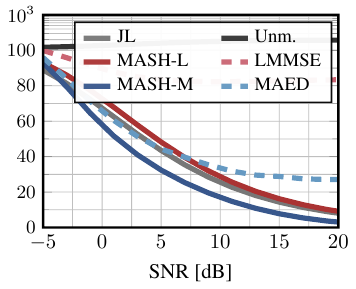}
}\!\!\!\!\!

\!\!\!\!
\subfigure[multi-ant. eigenbeamf. jammer \tinygraycircled{5}]{
\includegraphics[height=3.35cm]{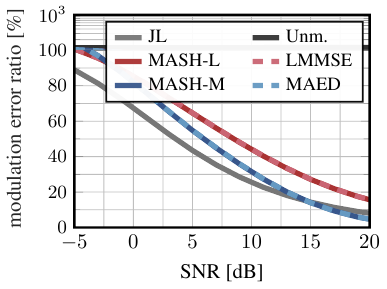}
}
\subfigure[multi-antenna data jammer \tinygraycircled{6}]{
\includegraphics[height=3.35cm]{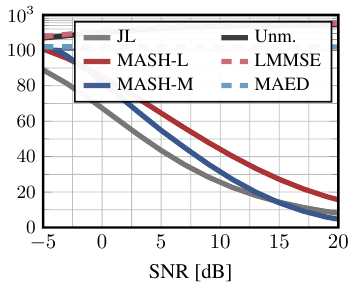}
}
\subfigure[multi-ant. dynamic jammer \tinygraycircled{7}]{
\includegraphics[height=3.35cm]{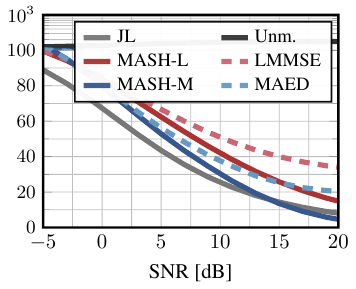}
}
\subfigure[multi-antenna repeat jammer \tinygraycircled{8}]{
\includegraphics[height=3.35cm]{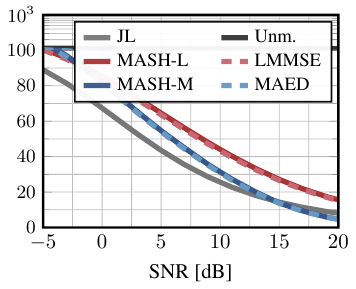}
}\!\!\!\!\!
\caption{
Modulation error ratio (MER) performance of different MASH receivers and baselines for eight types of jammers.
Note that the MER can exceed~$100\%$ when mitigation fails. Note also that 
the $y$-axis is scaled linearly between $0$ and $100$, and logarithmically between $100$ and $1000$.
\label{fig:mers}
\vspace{-2mm}
}
\end{figure*}

\subsection{Results} \label{sec:results}
The results in \fref{fig:jammers} and \fref{fig:mers} show that MASH mitigates \emph{all} jammers under consideration, 
and so confirm the universality of MASH. We start by discussing the BER results of \fref{fig:jammers}.

When facing the single-antenna barrage jammer in \fref{fig:barrage}, 
MASH-L and MASH-M have exactly the same performance as their non-MASH counterparts LMMSE and MAED. 
This is expected, since transforming a barrage jammer into its equivalent barrage jammer changes nothing.
LMMSE and~MASH-L perform close to the JL baseline.
MAED and \mbox{MASH-M} use a more complex, nonlinear JED-based data detector~\cite{marti2023maed}, and so are able to 
perform even better than the JL baseline.

For the more sophisticated single-antenna jammers, however, the picture changes (\fref{fig:pilot}--\fref{fig:sparse}): 
LMMSE performs poorly against all of them, since the training receive matrix~$\bY_J$ does not (or not~necessarily, 
in the case of the sparse jammer~\tinygraycircled{4}) contain samples in which the jammer is transmitting. 
MAED performs comparably bad.\footnote{In principle, MAED would be able to mitigate such jammers---\emph{if} 
given knowledge of $\Ie$ \cite{marti2023jmd, marti2024jmd}. Here, the problem is that MAED's estimate of $\Ie$ will (wrongly) be zero whenever $\bY_J$ 
contains no jammer samples.}
However, MASH-L and \mbox{MASH-M} transform all these jammers into their equivalent barrage~jammers and thus---as
predicted by theory---have \emph{exactly} the same performance as they do for the barrage jammer in \fref{fig:barrage}.

For the multi-antenna jammers, we observe the following (\fref{fig:eigenbeamforming}--\fref{fig:repeater}):
The eigenbeamforming jammer in \fref{fig:eigenbeamforming} is also a barrage jammer (even if it uses beamforming). 
So, as in \fref{fig:barrage}, the MASH methods have exactly the same performance
as their non-MASH counterparts. However, since the jammer now occupies 10 spatial dimensions (instead of~1) 
that need to be suppressed, the performance gap between the mitigation methods and the JL baseline (which does
not have to suppress any dimensions) is larger than in \fref{fig:barrage}
(the JMD methods MASH-M and MAED now perform worse than JL in spite of their nonlinear data detectors).
For the multi-antenna data jammers in \fref{fig:multi-data}, the non-MASH baselines fail spectacularly, 
while the MASH methods still have the same performance as for the multi-antenna barrage jammer in \fref{fig:eigenbeamforming}.
Similar obervations apply to the dynamic-beamforming jammer in \fref{fig:dynamic}, except that here, 
the non-MASH baselines can estimate parts of the interference subspace (or all of it, if they are lucky) 
during the training period and, thus, fare somewhat better. 
Finally, the repeat jammer in \fref{fig:repeater} behaves functionally almost like a barrage jammer and is sucessfully 
mitigated by all mitigation methods 
(note that the jammer transmits different signals for the MASH and the non-MASH methods). 
This experiment demonstrates that not even sneaky repeat attacks are able 
to overcome MASH. This was to be expected, of course, since $\text{row}(\Cpar)$ is, in general, not closed 
under cyclic shifts of the rows of $\Cpar$. 
(However, to prevent repeat attacks against MASH on the level of entire frames, the transmitters and receiver should 
update the matrix~$\bC$ according to a pseudo-random sequence after every transmission frame.)

The MER results in \fref{fig:mers} show strong agreement with the BER results in \fref{fig:jammers}.
The \emph{only} observable difference is that the nonlinear methods MAED/MASH-M perform slightly better compared the 
linear methods
JL/LMMSE/MASH-L in terms of MER than in terms of BER. (This is due to the data symbol 
prior of MAED/\mbox{MASH-M}, 
which pulls symbol estimates to constellation points and so achieves lower Euclidean error than linear detection
even when the underlying decoded bits are the same.) 
Even so, the results of \fref{fig:mers} agree with those of \fref{fig:jammers}: the MASH methods are the only 
ones that successfully mitigate \emph{all} types of jammers, while the traditional 
baselines that do not use subspace embeddings fail in many of the cases.

\begin{figure*}[tp]
\centering
\!\!\!\!\!\!
\subfigure[single-antenna barrage jammer \tinygraycircled{1}]{
\includegraphics[height=3.35cm]{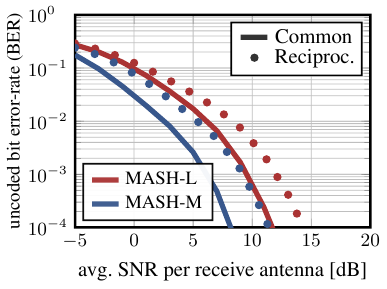}
\label{fig:recip:barrage}
}\!\!\!
\subfigure[single-antenna data jammer \tinygraycircled{2}]{
\includegraphics[height=3.35cm]{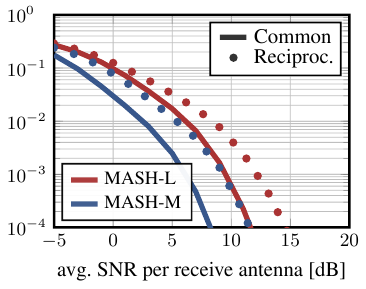}
\label{fig:recip:data}
}\!\!\!
\subfigure[single-antenna pilot jammer \tinygraycircled{3}]{
\includegraphics[height=3.35cm]{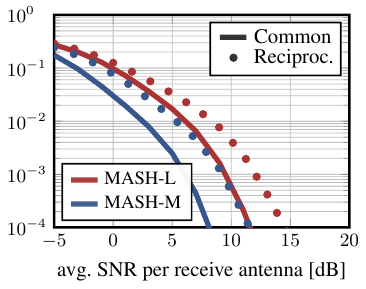}
\label{fig:recip:pilot}
}\!\!\!
\subfigure[single-antenna sparse jammer \tinygraycircled{4}]{
\includegraphics[height=3.35cm]{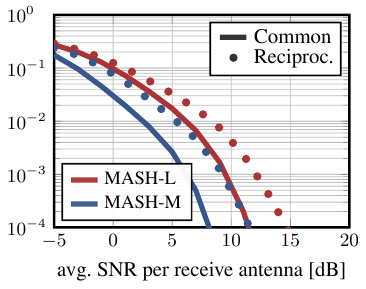}
\label{fig:recip:sparse}
}\!\!\!\!
\!\!\!\!\!\!
\subfigure[multi-ant. eigenbeamf. jammer \tinygraycircled{5}]{
\includegraphics[height=3.35cm]{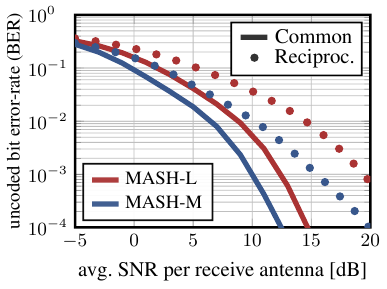}
\label{fig:recip:eigenbeamforming}
}\!\!\!
\subfigure[multi-antenna data jammer \tinygraycircled{6}]{
\includegraphics[height=3.35cm]{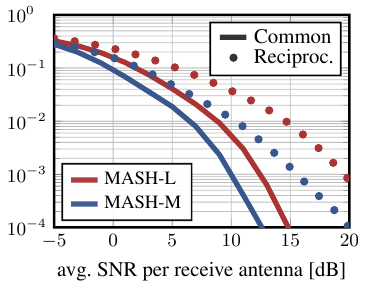}
\label{fig:recip:multi-data}
}\!\!\!
\subfigure[multi-ant. dynamic jammer \tinygraycircled{7}]{
\includegraphics[height=3.35cm]{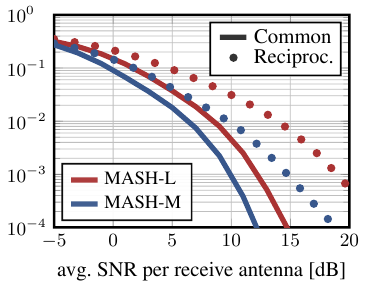}
\label{fig:recip:dynamic}
}\!\!\!
\subfigure[multi-antenna repeat jammer \tinygraycircled{8}]{
\includegraphics[height=3.35cm]{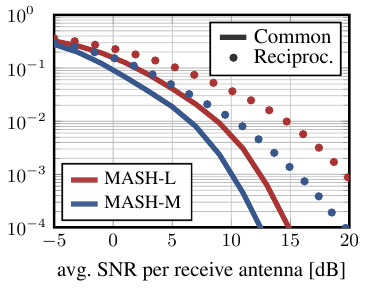}
\label{fig:recip:repeater}
}\!\!\!\!
\caption{
Bit error rate (BER) performance comparison between MASH with a common secret shared between the BS and all UEs (Common)
and reciprocal~MASH. 
}
\vspace{-2mm}
\label{fig:recip}
\end{figure*}

\subsection{Evaluation of Reciprocal MASH}
Next, we evaluate the performance of reciprocal MASH from \fref{sec:recip}. 
For this, the setup is exactly as in \fref{sec:setup}, except that we now consider only $U=8$ UEs---for 
reciprocal MASH, the performance decreases considerably as $U+\Ie$ approaches $R$. 
Since the performance of MASH vis-\`a-vis the state of the art has been evaluated 
in \fref{sec:results}, we now only compare reciprocal and non-reciprocal MASH. 
For reciprocal MASH, the following two receivers are considered:
\subsubsection*{\textbf{Recip. MASH-L}} 
This receiver uses reciprocal MASH as in \fref{sec:recip} with an LMMSE receiver as in \fref{sec:mash-lmmse}. 
\subsubsection*{\textbf{Recip. MASH-M}}
This receiver uses reciprocal MASH as in \fref{sec:recip} with an LMMSE receiver as in \fref{sec:mash-maed}. 
However, in contrast to MASH-M as described in \fref{sec:methods}, the reciprocal variant uses only $t_{\max}=5$ 
iterations per~\mbox{single-UE} detection problem, and it estimates the interference subspace dimension 
as the number of singular values (averaged over $u=1,\dots,U$) of $\bar{\bY}_{J,u}$ that exceed $\sqrt{B\No}$
(instead of $2\sqrt{B\No}$).\footnote{When, as in this experiment, all UEs are jointly synchronized, 
then the dimension of the interfernece subspace is the same for all UE detection problems. Hence, the
estimation of this dimension can be improved by considering the observations from all UEs jointly.}

Noting the close agreement between BER and MER, 
we evaluate only the BER to assess the performance. We still consider all types 
of jammers \tinygraycircled{1}\,--\,\tinygraycircled{8} (with jamming power $\rho=30$\,dB), 
but the multi-antenna jammers no have only $I=6$ antennas (to ensure that $R\geq I+U-1$ holds, since $R=16$).

The results are shown in \fref{fig:recip}. 
As mentioned in \fref{sec:recip}, treating treating the other $U-1$ UEs as additional interferers
is suboptimal for detection, and there is a corresponding performance loss for reciprocal MASH compared to 
regular (``common'') MASH. For the single-antenna jammers \mbox{(\fref{fig:recip:barrage}--\fref{fig:recip:sparse}),} 
reciprocal MASH-L and MASH-M lose roughly 3\,dB and 4\,dB in SNR (at a BER of $10^{-3}$) compared to their non-reciprocal counterparts, 
respectively. For the multi-antenna jammers (\fref{fig:recip:eigenbeamforming}--\fref{fig:recip:repeater}), these losses
become more pronounced since the dimension $\Ie+U-1$ of the ``interference'' subspace (for reciprocal MASH)
approaches the redundancy $R$. Note, however, that reciprocal MASH does not exhibit any signs of an error floor. 
The performance gap could be narrowed by increasing the redundancy $R$, 
at the expense of reduced number of data symbols per transmission frame.

\begin{figure*}[tp]
\centering
\!\!\!\!\!\!
\subfigure[single-antenna barrage jammer \tinygraycircled{1}]{
\includegraphics[height=3.35cm]{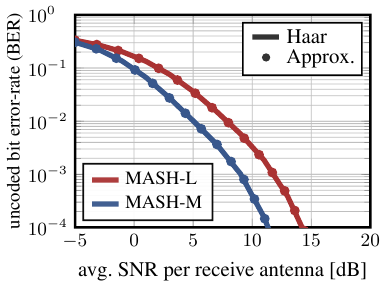}
\label{fig:approx:barrage}
}\!\!\!
\subfigure[single-antenna data jammer \tinygraycircled{2}]{
\includegraphics[height=3.35cm]{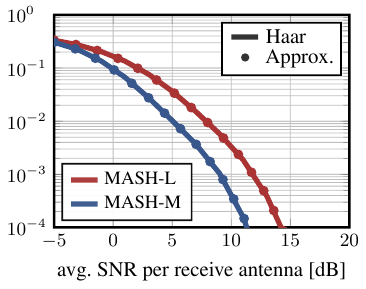}
\label{fig:approx:data}
}\!\!\!
\subfigure[single-antenna pilot jammer \tinygraycircled{3}]{
\includegraphics[height=3.35cm]{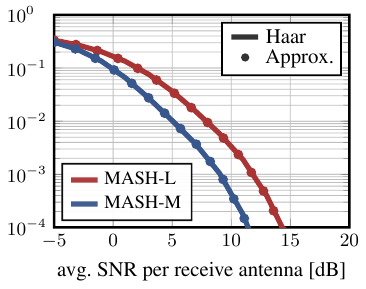}
\label{fig:approx:pilot}
}\!\!\!
\subfigure[single-antenna sparse jammer \tinygraycircled{4}]{
\includegraphics[height=3.35cm]{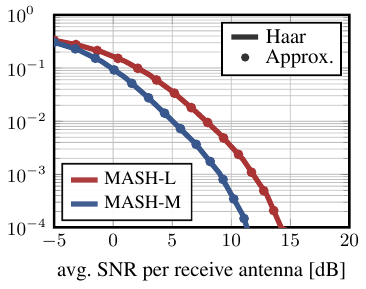}
\label{fig:approx:sparse}
}\!\!\!\!
\!\!\!\!\!\!
\subfigure[multi-ant. eigenbeamf. jammer \tinygraycircled{5}]{
\includegraphics[height=3.35cm]{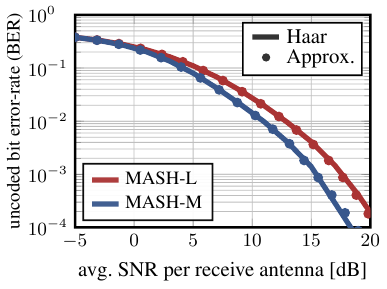}
\label{fig:approx:eigenbeamforming}
}\!\!\!
\subfigure[multi-antenna data jammer \tinygraycircled{6}]{
\includegraphics[height=3.35cm]{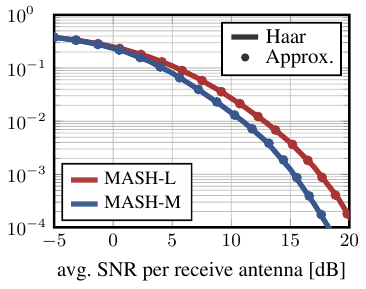}
\label{fig:approx:multi-data}
}\!\!\!
\subfigure[multi-ant. dynamic jammer \tinygraycircled{7}]{
\includegraphics[height=3.35cm]{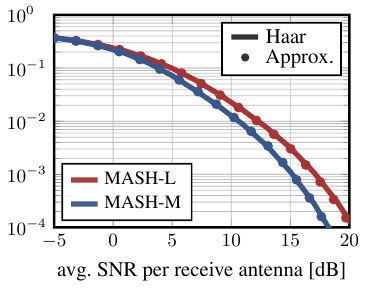}
\label{fig:approx:dynamic}
}\!\!\!
\subfigure[multi-antenna repeat jammer \tinygraycircled{8}]{
\includegraphics[height=3.35cm]{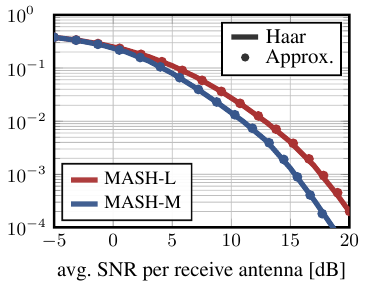}
\label{fig:approx:repeater}
}\!\!\!\!
\caption{
Bit error rate (BER) performance comparison between MASH with Haar distributed transform matrices 
and with approximately Haar distributed transform matrices (Approx.). 
}
\vspace{-2mm}
\label{fig:approx}
\end{figure*}

\subsection{Evaluation of Efficient Embedding/Raising Transforms}
Finally, we evaluate the performance of the efficient embedding/raising transforms from \fref{sec:efficient}, specifically
the version based on the Walsh-Hadamard matrix.  
The setup is again as in \fref{sec:setup}, except that the frame length is now $L=96$ (instead of $L=100$), 
resulting in $L-R-T=64$ data samples. The reason for this change is that there exists no Walsh-Hadamard matrix of 
size $100\times100$, but there exists one of size $96\times96$. We consider the following two new methods:

\subsubsection*{\textbf{Approx. MASH-L}} 
This method uses the Walsh-Hadamard-based matrix  from \fref{sec:efficient} for embedding and raising. 
The receiver operates like the MASH-L receiver from \fref{sec:methods}.
\subsubsection*{\textbf{Approx. MASH-M}}
This method uses the Walsh-Hadamard-based matrix  from \fref{sec:efficient} for embedding and raising. 
The receiver operates like the MASH-M receiver from \fref{sec:methods}.

We compare these two methods with their Haar distributed counterparts from \fref{sec:methods}. 
The BER results for all types of jammers \tinygraycircled{1}\,--\,\tinygraycircled{8} are depicted in \fref{fig:approx}.
For the single-antenna jammers, there is no discernible difference between MASH based on exact Haar matrices and 
MASH based on aproximate Haar matrices (\fref{fig:approx:barrage}--\fref{fig:approx:sparse}). 
Also for the multi-antenna jammers, the performance difference is less than $0.5$\,dB in all cases. 
This indicates that the efficient embedding/raising transforms succeed in approximating the behavior of 
Haar-based transform matrices for the purposes of MASH. 
Also note that, for barrage jammers, the performance is exactly identical, since barrage jammers are statistically
invariant to unitary transforms (which includes Haar matrices as well as the approximate transforms of 
\fref{sec:efficient}), see \fref{fig:approx:barrage},~\fref{fig:approx:eigenbeamforming}.

\section{Conclusions}
We have provided a mathematical definition to capture the essence of the notion of barrage jammers, 
which are easy to mitigate using multi-antenna spatial filtering. 
For single-antenna barrage jammers, we have also derived a fundamental new result in \fref{prop:residual}, 
which \emph{proves} that it is easy to mitigate them using MIMO processing. 
Furthermore, we have proposed MASH, a novel method in which
the transmitters embed their signals in a secret temporal subspace out of which the receiver raises them, 
thereby provably (cf. \fref{thm:barrage}) transforming \emph{any} jammer into a barrage jammer. 
Considering a massive MU-MIMO uplink example scenario, 
we have provided three concrete variations of MASH with different performance-complexity tradeoffs. 
Furthermore, we have proposed a way to remedy the difficulty of supplying a common secret to multiple UEs 
(while ensuring that the jammer does not learn about the secret) by proposing \emph{reciprocal} MASH. 
In reciprocal MASH, each UE uses its own secret (known only to the BS and that UE), while the BS 
detects the data of the different UEs and treats the respective other UEs as additional interference.
We then have suggested the use of transform matrices that are approximately Haar distributed, 
but that reduce both the computational complexity of the legitimate transceivers as well as the required
entropy of the shared random secret. 
Finally, the efficacy of MASH in all its variants, and against many conceivable types of jammers, has been 
demonstrated using extensive numerical simulation results. 

\appendices
\section{Proof of \fref{prop:residual}} \label{app:proof1}
The proof consists of two cases that are considered. In one case, we assume that the noise $\bN$ satisfies
\begin{align}\textstyle
	\|\bN_{[1:R]}\|^2 \leq \alpha^{-1}\left(\frac32 - \sqrt{2}\right)\|\bmj\|_2^2\|\bmw\|_2^2 \frac{L}{R}, \label{eq:noise_bound}
\end{align}
and in the other case, we assume the reverse. 
We start with the case where \eqref{eq:noise_bound} holds. 
The compact SVD of the interference $\bmj\tp{\bmw}$ is $\bsfu\sfsigma\herm{\bsfv}$, 
with $\sfsigma=\|\bmj\|_2\|\bmw\|_2$, and with $\bsfv$ being uniformly distributed 
over the complex $L$-dimensional unit sphere (since the jammer is a barrage jammer).  
We may therefore  write
\begin{align}
	(\bmj\tp{\bmw}+\bN)_{[1:R]} &= \bsfu\sfsigma(\herm{\bsfv})_{[1:R]} + \bN_{[1:R]} \\
	&= \bsfu (\sfsigma\|\bsfv_{[1:R]}\|_2)\frac{(\herm{\bsfv})_{[1:R]}}{\|\bsfv_{[1:R]}\|_2} + \bN_{[1:R]}.
\end{align}
From \cite[Lem. 2.2]{dasgupta2003elementary}, it follows that, for any $\alpha>1$, 
\begin{align}\textstyle
	\|\bsfv_{[1:R]}\|_2 >  \sqrt{\frac{R}{\alpha L}} 
\end{align}
holds with probability at least 
\begin{align}\textstyle
	1-\alpha^{-R} \left( \frac{L-R/\alpha}{L-R}\right)^{L-R}. \label{eq:probab_bound}
\end{align}
For the remainder of this first case, we condition on this event.
Wedin's $\sin\Theta$ theorem \cite[Thm. 2.9]{chen2021spectral} implies that if
$\|\bN_{[1:R]}\|\leq \left(1 - \frac{1}{\sqrt2}\right)\sigma\|\bsfv_{[1:R]}\|_2$---which 
is implied by \eqref{eq:noise_bound}---then the estimate $\bmu_1$ of the jammer subspace $\bsfu$ satisfies
\begin{align}
\|\bmu_1\herm{\bmu_1} - \bsfu\herm{\bsfu}\| \leq \frac{2\|\bN_{[1:R]}\|}{\sigma\|\bsfv_{[1:R]}\|_2} 
< \frac{2\|\bN_{[1:R]}\|}{\|\bmj\|_2\|\bmw\|_2 \sqrt{R/(\alpha L)}}.\! \label{eq:proj_bound}
\end{align}
We can therefore bound the square root of the residual jammer interference energy after the projection as 
\begin{align}
	&\|\bP\bmj\tp{\bmw}\|_\textnormal{F} = \|(\bI_B-\bmu_1\herm{\bmu_1})\bmj\|_2 \|\bmw\|_2 \\
	&\hphantom{\|\bP\bmj\tp{\bmw}\|_\textnormal{F}}
	= \|(\bI_B - \bsfu\herm{\bsfu} + \bsfu\herm{\bsfu} -\bmu_1\herm{\bmu_1})\bmj\|_2 \|\bmw\|_2 \\
	&\leq \underbrace{\|(\bI_B - \bsfu\herm{\bsfu}) \bmj\|_2}_{=0} \|\bmw\|_2 
	+ \|(\bsfu\herm{\bsfu} -\bmu_1\herm{\bmu_1})\bmj\|_2 \|\bmw\|_2 \label{eq:triangle_ineq} \\
	&\leq \|\bmu_1\herm{\bmu_1} - \bsfu\herm{\bsfu}\| \|\bmj\|_2 \|\bmw\|_2 \\
	&\leq \frac{2\|\bN_{[1:R]}\|}{\|\bmj\|_2\|\bmw\|_2 \sqrt{R/(\alpha L)}} \|\bmj\|_2 \|\bmw\|_2 \label{eq:use_proj_bound} \\
	&= 2\|\bN_{[1:R]}\| \sqrt{\frac{\alpha L}{R}},
\end{align}
where in \eqref{eq:triangle_ineq} we used the triangle inequality and the fact that $(\bI_B - \bsfu\herm{\bsfu}) \bmj=\mathbf{0}$
since $\bI_B - \bsfu\herm{\bsfu}$ is the projection onto the orthogonal complement of $\text{col}(\bmj)$, 
and where in \eqref{eq:use_proj_bound} we used~\eqref{eq:proj_bound}.
By taking the square, we therefore get
\begin{align}
	\|\bP\bmj\tp{\bmw}\|_\textnormal{F}^2 \leq \alpha 4 \|\bN_{[1:R]}\|^2 \frac{L}{R} \label{eq:energy_bound_1}
\end{align}

In the second case, we assume that \eqref{eq:noise_bound} does \emph{not} hold. 
In that case, we can bound the residual jammer interference energy as
\begin{align}
	\|\bP\bmj\tp{\bmw}\|_\textnormal{F}^2 &\leq \|\bmj\tp{\bmw}\|_\textnormal{F}^2 = \|\bmj\|_2^2 \|\bmw\|_2^2 \label{eq:contraction} \\
	&< \alpha\frac{1}{\frac32-\sqrt{2}}\|\bN_{[1:R]}\|^2 \frac{L}{R} \label{eq:energy_bound_2},
\end{align}
where in \eqref{eq:contraction} we use the fact that projections are contractions, 
and where in \eqref{eq:energy_bound_2} we use that \eqref{eq:noise_bound} does not hold. 

In summary, we get that if \eqref{eq:noise_bound} holds, then \eqref{eq:energy_bound_1} holds with probability 
at least \eqref{eq:probab_bound};  
and if \eqref{eq:noise_bound} does not hold, then \eqref{eq:energy_bound_2} holds with probability 1. 
By taking the weaker bounds of both cases, it follows that 
\begin{align}
	\|\bP\bmj\tp{\bmw}\|_\textnormal{F}^2 &\leq \alpha\frac{1}{\frac32-\sqrt{2}}\|\bN_{[1:R]}\|^2 \frac{L}{R} 
\end{align}
holds with probability at least \eqref{eq:probab_bound}, regardless of whether \eqref{eq:noise_bound} holds. 
The result follows dividing both sides by $L$, and by using the facts $\|\bN_{[1:R]}\|^2\leq\|\bN_{[1:R]}\|_\textnormal{F}^2$ and 
$\frac{1}{\frac32-\sqrt{2}}<12$. 
$\hfill\largesecret$
\vfill

\section{Proof of \fref{thm:barrage}} \label{app:app}
We rewrite the compact SVD of $\bJ\bW = \bsfU\bsfSigma\herm{\bsfV}$ as a full SVD
$\bJ\bW = \bU\bSigma\herm{\bV}$ with $\bU\in\opC^{B\times B}, \bSigma\in\opC^{B\times L},$ and $\bV\in\opC^{L\times L}$. 
We then have $\bJ\bW\herm{\bC} = \bU\bSigma\herm{(\bC\bV)}$.
Since~$\bC$ is Haar distributed, so is $\bC\bV$ \cite[Thm.\,1.4]{meckes2019random}. 
The compact SVD of $\bJ\bW\herm{\bC}\!=\!\bar{\bsfU}\bar{\bsfSigma}\herm{\bar{\bsfV}}$ is 
$\bar{\bsfU} \!=\! \bU_{[1:\Ie]}=\bsfU$, \mbox{$\bar{\bsfSigma}\!=\!\bSigma_{(1:\Ie),[1:\Ie]}=\bsfSigma$,}
which implies $\bar{\bsigma}=\bsigma$, and 
$\bar{\bsfV} = (\bC\bV)_{[1:\Ie]}$. Since $\bC\bV$ is Haar distributed, the columns of $\bar{\bsfV}$ are
distributed uniformly over the complex $L$-dimensional sphere. $\hfill\largesecret$

\balance


\begin{thebibliography}{10}
\providecommand{\url}[1]{#1}
\csname url@samestyle\endcsname
\providecommand{\newblock}{\relax}
\providecommand{\bibinfo}[2]{#2}
\providecommand{\BIBentrySTDinterwordspacing}{\spaceskip=0pt\relax}
\providecommand{\BIBentryALTinterwordstretchfactor}{4}
\providecommand{\BIBentryALTinterwordspacing}{\spaceskip=\fontdimen2\font plus
\BIBentryALTinterwordstretchfactor\fontdimen3\font minus
  \fontdimen4\font\relax}
\providecommand{\BIBforeignlanguage}[2]{{%
\expandafter\ifx\csname l@#1\endcsname\relax
\typeout{** WARNING: IEEEtran.bst: No hyphenation pattern has been}%
\typeout{** loaded for the language `#1'. Using the pattern for}%
\typeout{** the default language instead.}%
\else
\language=\csname l@#1\endcsname
\fi
#2}}
\providecommand{\BIBdecl}{\relax}
\BIBdecl

\bibitem{marti2023universal}
G.~Marti and C.~Studer, ``Universal {MIMO} jammer mitigation via secret
  temporal subspace embeddings,'' pp. 1--8, Oct 2023.

\bibitem{dahlman20205g}
E.~Dahlman, S.~Parkvall, and J.~Skold, \emph{5G NR: The next generation
  wireless access technology}.\hskip 1em plus 0.5em minus 0.4em\relax Academic
  Press, 2020.

\bibitem{lichtman20185g}
M.~Lichtman, R.~Rao, V.~Marojevic, J.~Reed, and R.~P. Jover, ``{5G NR} jamming,
  spoofing, and sniffing: Threat assessment and mitigation,'' in \emph{Proc.
  IEEE Int. Conf. Commun. Workshop (ICCW)}, May 2018, pp. 1--6.

\bibitem{girke2019towards}
F.~Girke, F.~Kurtz, N.~Dorsch, and C.~Wietfeld, ``Towards resilient {5G}:
  Lessons learned from experimental evaluations of {LTE} uplink jamming,'' in
  \emph{IEEE Int. Conf. Commun. Workshop (ICCW)}, May 2019, pp. 1--6.

\bibitem{threatvectors2021cisa}
{5G Threat Model Working Panel}, ``Potential threat vectors to {5G}
  infrastructure,'' \emph{CISA, NSA, and DNI}, May 2021.

\bibitem{guanella1944dsss}
G.~Guanella, ``Means for and method of secret signaling,'' U.S. Patent 2405500,
  Aug. 1946.

\bibitem{madhow1994mmse}
U.~Madhow and M.~L. Honig, ``{MMSE} interference suppression for
  direct-sequence spread-spectrum {CDMA},'' \emph{{IEEE} Trans. Commun.},
  vol.~42, no.~12, pp. 3178--3188, Dec. 1994.

\bibitem{tesla1903fhss}
N.~Tesla, ``System of signaling,'' U.S. Patent 725605A, Apr. 1903.

\bibitem{stark1985coding}
W.~Stark, ``Coding for frequency-hopped spread-spectrum communication with
  partial-band interference-part i: capacity and cutoff rate,'' vol.~33,
  no.~10, pp. 1036--1044, 1985.

\bibitem{leost2012interference}
Y.~L{\'e}ost, M.~Abdi, R.~Richter, and M.~Jeschke, ``Interference rejection
  combining in {LTE} networks,'' \emph{Bell Labs Tech.~J.}, vol.~17, no.~1, pp.
  25--50, Jun. 2012.

\bibitem{miller2010subverting}
R.~Miller and W.~Trappe, ``Subverting {MIMO} wireless systems by jamming the
  channel estimation procedure,'' in \emph{Proc. ACM Conf. Wireless Netw.
  Security}, Mar. 2010, pp. 19--24.

\bibitem{lichtman2016communications}
M.~Lichtman, J.~D. Poston, S.~Amuru, C.~Shahriar, T.~C. Clancy, R.~M. Buehrer,
  and J.~H. Reed, ``A communications jamming taxonomy,'' \emph{IEEE Security \&
  Privacy}, vol.~14, no.~1, pp. 47--54, 2016.

\bibitem{pirayesh2022jamming}
H.~Pirayesh and H.~Zeng, ``Jamming attacks and anti-jamming strategies in
  wireless networks: A comprehensive survey,'' \emph{{IEEE} Commun. Surveys
  Tuts.}, vol.~9, no.~2, pp. 767--809, 2022.

\bibitem{marti2021hybrid}
G.~Marti, O.~Casta\~neda, S.~Jacobsson, G.~Durisi, T.~Goldstein, and C.~Studer,
  ``Hybrid jammer mitigation for all-digital {mmWave} massive {MU-MIMO},'' in
  \emph{Proc. Asilomar Conf. Signals, Syst., Comput.}, Nov. 2021, pp. 93--99.

\bibitem{zhu2019mitigating}
J.~Zhu, Z.~Wang, Q.~Li, H.~Chen, and N.~Ansari, ``Mitigating intended jamming
  in {mmWave MIMO} by hybrid beamforming,'' \emph{{IEEE} Wireless Commun.
  Lett.}, vol.~8, no.~6, pp. 1617--1620, 2019.

\bibitem{jiang2021efficient}
X.~Jiang, X.~Liu, R.~Chen, Y.~Wang, F.~Shu, and J.~Wang, ``Efficient receive
  beamformers for secure spatial modulation against a malicious full-duplex
  attacker with eavesdropping ability,'' \emph{{IEEE} Trans. Veh. Technol.},
  vol.~70, no.~2, pp. 1962--1966, 2021.

\bibitem{chehimi2023machine}
M.~Chehimi, M.~K. Awad, M.~Al-Husseini, and A.~Chehab, ``Machine learning-based
  anti-jamming technique at the physical layer,'' \emph{Concurrency and
  Computation: Practice and Experience}, vol.~35, no.~9, 2023.

\bibitem{marti2021snips}
G.~Marti, O.~Casta\~neda, and C.~Studer, ``Jammer mitigation via beam-slicing
  for low-resolution {mmWave} massive {MU-MIMO},'' \emph{{IEEE} Open J.
  Circuits Syst.}, vol.~2, pp. 820--832, 2021.

\bibitem{yang2022estimation}
L.~Yang, X.~Jiang, F.~Shu, W.~Zhang, and J.~Wang, ``Estimation of covariance
  matrix of interference for secure spatial modulation against a malicious
  full-duplex attacker,'' \emph{{IEEE} Trans. Veh. Technol.}, vol.~71, no.~8,
  pp. 9050--9054, Aug. 2022.

\bibitem{he2022high}
H.~He, T.~Su, H.~Wang, Y.~Teng, W.~Shi, F.~Shu, and J.~Wang, ``High-performance
  estimation of jamming covariance matrix for {IRS}-aided directional
  modulation network with a malicious attacker,'' \emph{{IEEE} Trans. Veh.
  Technol.}, vol.~71, no.~9, pp. 10\,137--10\,142, Sep. 2022.

\bibitem{do18a}
T.~T. {Do}, E.~{Bj\"ornsson}, E.~G. {Larsson}, and S.~M. {Razavizadeh},
  ``Jamming-resistant receivers for the massive {MIMO} uplink,'' \emph{{IEEE}
  Trans. Inf. Forensics Security}, vol.~13, no.~1, pp. 210--223, Jan. 2018.

\bibitem{akhlaghpasand20a}
H.~{Akhlaghpasand}, E.~{Bj\"ornsson}, and S.~M. {Razavizadeh}, ``Jamming
  suppression in massive {MIMO} systems,'' \emph{{IEEE} Trans. Circuits Syst.
  {II}}, vol.~68, no.~1, pp. 182--186, Jan. 2020.

\bibitem{akhlaghpasand20b}
H.~{Akhlaghpasand}, E.~{Bj\"ornsson}, and S.~{Razavizadeh}, ``{Jamming-robust}
  uplink transmission for spatially correlated massive {MIMO} systems,''
  \emph{{IEEE} Trans. Commun.}, vol.~68, no.~6, pp. 3495--3504, Jun. 2020.

\bibitem{zeng2017enabling}
H.~Zeng, C.~Cao, H.~Li, and Q.~Yan, ``Enabling jamming-resistant communications
  in wireless {MIMO} networks,'' in \emph{Proc. IEEE Conf. Commun. Netw.
  Security (CNS)}, Oct. 2017, pp. 1--9.

\bibitem{shen14a}
W.~{Shen}, P.~{Ning}, X.~{He}, H.~{Dai}, and Y.~{Liu}, ``{MCR} decoding: A
  {MIMO} approach for defending against wireless jamming attacks,'' in
  \emph{Proc. IEEE Conf. Commun. Netw. Security (CNS)}, Oct. 2014, pp.
  133--138.

\bibitem{yan2016jamming}
Q.~Yan, H.~Zeng, T.~Jiang, M.~Li, W.~Lou, and Y.~T. Hou, ``Jamming resilient
  communication using {MIMO} interference cancellation,'' \emph{{IEEE} Trans.
  Inf. Forensics Security}, vol.~11, no.~7, pp. 1486--1499, Jul. 2016.

\bibitem{hoang2021suppression}
L.~M. Hoang, J.~A. Zhang, D.~N. Nguyen, X.~Huang, A.~Kekirigoda, and K.-P. Hui,
  ``Suppression of multiple spatially correlated jammers,'' \emph{{IEEE} Trans.
  Veh. Technol.}, vol.~70, no.~10, pp. 10\,489--10\,500, 2021.

\bibitem{hoang2022multiple}
L.~M. Hoang, D.~Nguyen, J.~A. Zhang, and D.~T. Hoang, ``Multiple correlated
  jammers suppression: A deep dueling {Q}-learning approach,'' in \emph{Proc.
  IEEE Wireless Commun. Netw. Conf. (WCNC)}, Apr. 2022, pp. 998--1003.

\bibitem{marti2023maed}
G.~Marti, T.~K\"olle, and C.~Studer, ``Mitigating smart jammers in multi-user
  {MIMO},'' \emph{{IEEE} Trans. Signal Process.}, vol.~71, pp. 756--771, 2023.

\bibitem{marti2023jmd}
G.~Marti and C.~Studer, ``Joint jammer mitigation and data detection for smart,
  distributed, and multi-antenna jammers,'' in \emph{Proc. IEEE Int. Conf.
  Commun. (ICC)}, May 2023, pp. 1--6.

\bibitem{marti2023single}
G.~Marti and C.~Studer, ``Single-antenna jammers in {MIMO-OFDM} can resemble
  multi-antenna jammers,'' \emph{{IEEE} Commun. Lett.}, vol.~27, no.~11, pp.
  3103--3107, Nov. 2023.

\bibitem{meckes2019random}
E.~S. Meckes, \emph{The random matrix theory of the classical compact
  groups}.\hskip 1em plus 0.5em minus 0.4em\relax Cambridge University Press,
  2019, vol. 218.

\bibitem{eckart1936approximation}
C.~Eckart and G.~Young, ``The approximation of one matrix by another of lower
  rank,'' \emph{Psychometrika}, vol.~1, no.~3, pp. 211--218, 1936.

\bibitem{liavas2001behavior}
A.~P. Liavas and P.~A. Regalia, ``On the behavior of information theoretic
  criteria for model order selection,'' \emph{{IEEE} Trans. Signal Process.},
  vol.~49, no.~8, pp. 1689--1695, 2001.

\bibitem{petersen12a}
K.~B. Petersen and M.~S. Pedersen, ``The matrix cookbook,'' Nov. 2012.

\bibitem{marti2024jmd}
G.~Marti and C.~Studer, ``Joint jammer mitigation and data detection,'' in
  preparation.

\bibitem{barzilai1988two}
J.~Barzilai and J.~M. Borwein, ``Two-point step size gradient methods,''
  \emph{IMA J. Numer. Anal.}, vol.~8, no.~1, pp. 141--148, 1988.

\bibitem{goldstein16a}
\BIBentryALTinterwordspacing
T.~Goldstein, C.~Studer, and R.~G. Baraniuk, ``A field guide to
  forward-backward splitting with a {FASTA} implementation,'' Feb. 2016.
  [Online]. Available: \url{https://arxiv.org/abs/1411.3406}
\BIBentrySTDinterwordspacing

\bibitem{jover2014enhancing}
R.~P. Jover, J.~Lackey, and A.~Raghavan, ``Enhancing the security of {LTE}
  networks against jamming attacks,'' \emph{EURASIP J. Inf. Security}, vol.
  2014, no.~1, pp. 1--14, 2014.

\bibitem{omri2019synchronization}
A.~Omri, M.~Shaqfeh, A.~Ali, and H.~Alnuweiri, ``Synchronization procedure in
  {5G NR} systems,'' \emph{IEEE Access}, vol.~7, pp. 41\,286--41\,295, 2019.

\bibitem{marti2024jass}
G.~Marti, F.~Arquint, and C.~Studer, ``Jammer-resilient time-synchronization in
  the {MIMO} uplink,'' \emph{arXiv preprint arXiv:2404.05335}.

\bibitem{ailon2014fast}
N.~Ailon and H.~Rauhut, ``Fast and rip-optimal transforms,'' \emph{Discrete \&
  Computational Geometry}, vol.~52, no.~4, pp. 780--798, 2014.

\bibitem{andoni2015practical}
A.~Andoni, P.~Indyk, T.~Laarhoven, I.~Razenshteyn, and L.~Schmidt, ``Practical
  and optimal lsh for angular distance,'' \emph{Advances in neural information
  processing systems}, vol.~28, 2015.

\bibitem{borel1914introduction}
E.~Borel, \emph{Introduction g{\'e}om{\'e}trique {\`a} quelques th{\'e}ories
  physiques}.\hskip 1em plus 0.5em minus 0.4em\relax Gauthier-Villars, 1914.

\bibitem{sutton2005stochastic}
B.~D. Sutton, ``The stochastic operator approach to random matrix theory,''
  Ph.D. dissertation, Massachusetts Institute of Technology, 2005.

\bibitem{jaeckel2014quadriga}
S.~Jaeckel, L.~Raschkowski, K.~B{\"o}rner, and L.~Thiele, ``{QuaDRiGa}: A {3-D}
  multi-cell channel model with time evolution for enabling virtual field
  trials,'' \emph{{IEEE} Trans. Antennas Propag.}, vol.~62, no.~6, pp.
  3242--3256, Jun. 2014.

\bibitem{3gpp22a}
3GPP, ``{3GPP TR 38.901},'' Mar. 2022, version 17.0.0.

\bibitem{3gpp21a}
3GPP, ``{5G}; {NR}; base station ({BS}) radio transmission and reception,''
  Mar. 2021, {TS} 38.104 version 17.1.0 Rel.~17.

\bibitem{dasgupta2003elementary}
S.~Dasgupta and A.~Gupta, ``An elementary proof of a theorem of johnson and
  lindenstrauss,'' \emph{Random Structures \& Algorithms}, vol.~22, no.~1, pp.
  60--65, 2003.

\bibitem{chen2021spectral}
Y.~Chen, Y.~Chi, J.~Fan, C.~Ma \emph{et~al.}, ``Spectral methods for data
  science: A statistical perspective,'' \emph{Foundations and
  Trends{\textregistered} in Machine Learning}, vol.~14, no.~5, pp. 566--806,
  2021.

\end{thebibliography}
\end{document}